%% file: scnet.tex
\documentclass[journal]{IEEEtran}

\usepackage[para]{threeparttable}
\usepackage{caption}
\usepackage[font=small]{caption}
\usepackage{subcaption}
\usepackage{multirow}
\usepackage{verbatim}
\usepackage{mymacros}
\usepackage{verbatim}

\begin{document}

\title{S\&CNet: Monocular Depth Completion for Autonomous Systems and 3D Reconstruction}

\begin{comment}
\author{Michael~Shell,~\IEEEmembership{Member,~IEEE,}
John~Doe,~\IEEEmembership{Fellow,~OSA,}
and~Jane~Doe,~\IEEEmembership{Life~Fellow,~IEEE}% <-this % stops a space
\thanks{M. Shell was with the Department
of Electrical and Computer Engineering, Georgia Institute of Technology, Atlanta,
GA, 30332 USA e-mail: (see http://www.michaelshell.org/contact.html).}% <-this % stops a space
\thanks{J. Doe and J. Doe are with Anonymous University.}% <-this % stops a space
\thanks{Manuscript received April 19, 2005; revised August 26, 2015.}}
\end{comment}

\author{
\thanks{This work is supported by National Natural Science Foundation of China under Grant 61620106012 and 61573048. Corresponding author: Weihai Chen and Zhengguo Li.}
%\thanks{$^*$ indicates the corresponding author of this paper.}
%\thanks{This work is supported by the National Natural Science Foundation of China (No.)}
\thanks{Lei Zhang, Weihai Chen and Xingming Wu are with the School of Automation Science and Electrical Engineering, Beihang University, 100191, Beijing, China (Emails: {leizhangbuaa@163.com,whchenbuaa@126.com})}
\thanks{Chao Hu is with the Peking University, Beijing, China. (Email: {1701210857@pku.edu.cn})}
\thanks{Zhengguo Li is with the Institute for Infocomm Research, 138632, Singapore (Email: {ezgli@i2r.a-star.edu.sg})}
%\textbf{Project website: \url{http://fastdepth.mit.edu}}}
Lei Zhang, Weihai Chen, Chao Hu, Xingming Wu, Zhengguo Li
}


\maketitle

\begin{abstract}
Dense depth completion is essential for autonomous systems and 3D reconstruction. In this paper, a lightweight yet efficient network (S\&CNet) is proposed to obtain a good trade-off between efficiency and accuracy for the dense depth completion. A dual-stream attention module (S\&C enhancer) is introduced to measure both spatial-wise and the channel-wise global-range relationship of extracted features so as to improve the performance. A coarse-to-fine network is designed and the proposed S\&C enhancer is plugged into the coarse estimation network between its encoder and decoder network. Experimental results demonstrate that our approach achieves competitive performance with existing works on KITTI dataset but almost four times faster. The proposed S\&C enhancer can be plugged into other existing works and boost their performance significantly with a negligible additional computational cost.
\end{abstract}


\IEEEpeerreviewmaketitle

\section{Introduction}
\begin{figure}[t] \centering
	\begin{subfigure}[b]{0.49\linewidth}
		\centering
		\includegraphics[width=45mm]{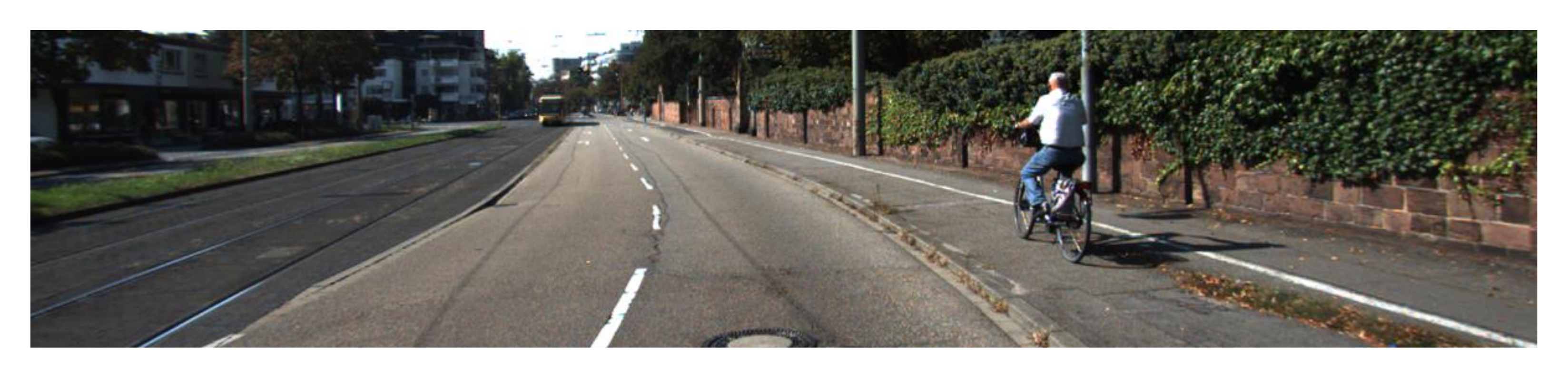}
		\caption{raw RGB image}
		\label{fig:a}
	\end{subfigure} %
	\begin{subfigure}[b]{0.49\linewidth}  
		\centering  
		\includegraphics[width=45mm]{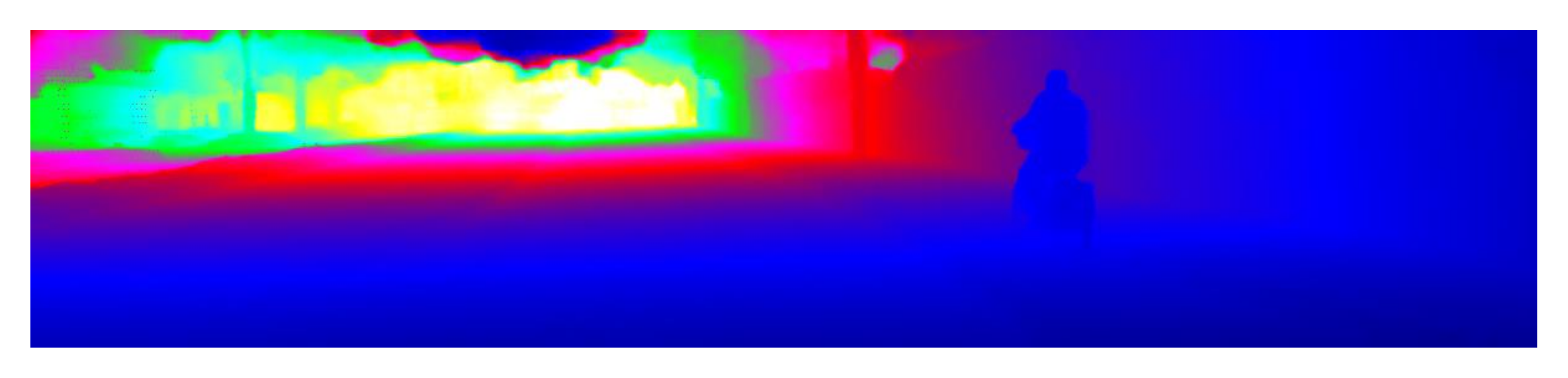}
		\caption{generated dense depth map}
		\label{fig:d} 
	\end{subfigure} 
	
	\begin{subfigure}[b]{0.49\linewidth}  
		\centering  
		\includegraphics[width=45mm]{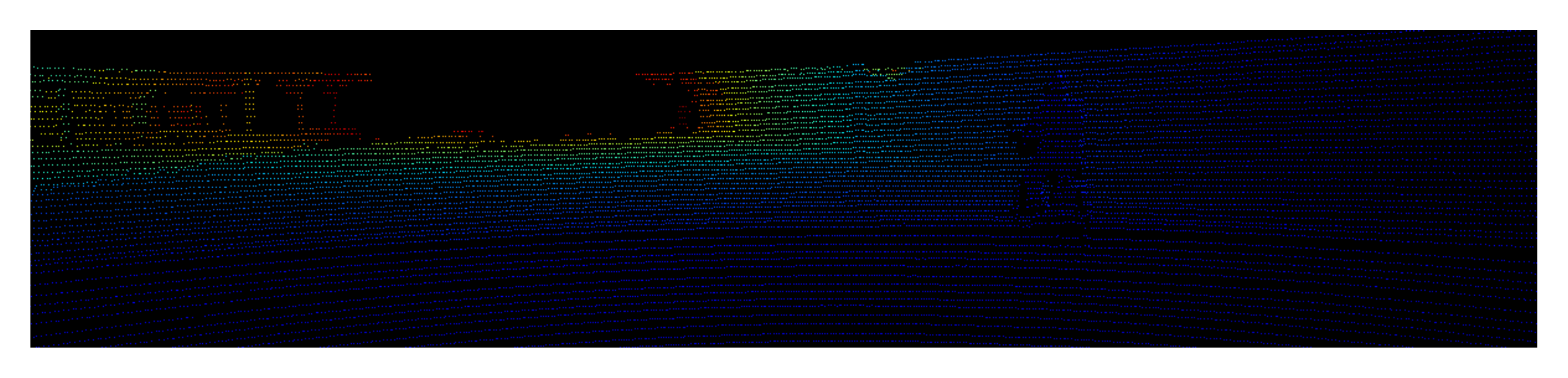}
		\caption{sparse depth map}  
		\label{fig:b}
	\end{subfigure} 
	\begin{subfigure}[b]{0.49\linewidth}  
		\centering  
		\includegraphics[width=45mm]{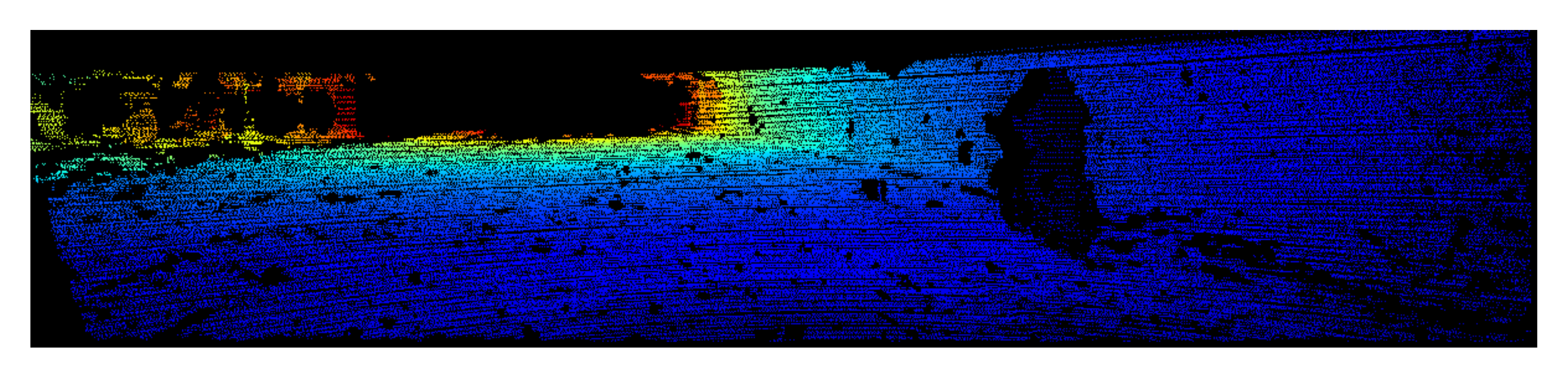}
		\caption{semi-dense depth map}  
		\label{fig:b}
	\end{subfigure}
	
	\caption{We develop a network for depth completion: given a raw rgb color image (a), and a sparse depth map (c), estimated a dense depth image (b). The semi-dense depth map (d) is utilized as groundtruth for training.}
	\label{fig:velodyne}
\end{figure}

\begin{comment}
\IEEEPARstart{T}{his} demo file is intended to serve as a ``starter file''
for IEEE journal papers produced under \LaTeX\ using
IEEEtran.cls version 1.8b and later.
I wish you the best of success.
\end{comment}
\label{sec:introduction}
Dense depth sensing is essential for robotics, autonomous vehicles and computer vision tasks, including obstacle avoidance, mapping and 3D reconstruction. Unfortunately, most depth sensors can only yield sparse measurement. As shown in Figure \ref{fig:b}, the 64-line Velodyne LiDAR scan can only provide sparse depth measurement. Therefore, the dense depth completion methods which estimate a dense depth map (e.g. Figure \ref{fig:d}) from sparse measurement (e.g. Figure \ref{fig:b}) produced by LiDARs or other sensors and its correlated RGB image  (e.g. Figure \ref{fig:a}) has been a growing interest task of industry as sparse yet high-accuracy depth map can be measured from sensors.

For the dense depth completion, current approaches have significant improvement in accuracy via encoder-decoder networks \cite{qiu2018deeplidar:}. Most of them achieve it at the cost of increasing computational complexity \cite{ma2018self-supervised} or introducing intermediate supervisors \cite{qiu2018deeplidar:}, which result in a slow inference speed. However, most real-time systems are not only limited by computational resources but also subject to latency constraints. Therefore, a lightweight yet efficient approach is essential for most of real-time systems. So far, many prior studies work on designing fast and efficient networks for image classification \cite{howard2017mobilenets:,sandler2018mobilenetv2:,hluchyj1991shuffle}, segmentation \cite{romera2018erfnet:} and object detection \cite{liu2016ssd:,redmon2016you,redmon2018v3:}. Among these works, the semantic segmentation is most similar to the dense depth completion as both of them can be considered as pixel-to-pixel translation problem.

Generally, sharp boundaries are essential for high-accuracy pixel-to-pixel translation problems. Many works show that a lower output stride (OS) of encoder network can preserve more detail information \cite{romera2018erfnet:,fu2018dual,chen2018encoder}. On the other hand, a lower output stride could limit the receptive filed of the backbone encoder network. Since a lightweight backbone in efficient network has already suffering from the low representational power, the limited receptive filed could further worsen the condition. Current approaches for solving this problem are mainly relying on the development of enhancer module \cite{yu2016multi-scale, zhao2017pyramid, chen2017rethinking, wang2018non-local}. Most recently, the non-local operation \cite{wang2018non-local} has been demonstrated that it can boost the performance of a backbone network with a global receptive filed and negligible additional computational cost. It can be expected that a non-local operation could improve the performance of low-OS encoder network. However, the original non-local operation put much effort on reducing the computational complexity by compressing the channels of identity features, which is efficient but may cause information loss. 

%%%%%%%%%%%%%%%%%%%%%%%%%%%%%%%%%%%%%%%%%%%%%%%%
%%%%%%%%%%%%%%%%%%%%%%%%%%%%%%%%%%%%%%%%%%%%%%%%

\begin{comment}
	Specifically, we found that each channel in the features generated by our encoder network responds to different distance. Intuitively, different channels should be paid different attention. For example, the decoder network should pay more attention on the channels response to close distance if one frame contains only near objects. Thus, it can be except that a channel-wise attention operation should also benefit the performance. Meanwhile, recent efforts tried on channel-wise attention module \cite{fu2018dual, hu2018squeeze-and-excitation, kim2018ram:} have demonstrated their great boost power in image semantic segmentation, classification and super-resolution tasks. Among these works, SE module \cite{ hu2018squeeze-and-excitation} is an efficient approach which utilize global descriptor of channels to regress weights of channels. However, the descriptor in SE module is simply measured by global average pooling, which may loss the high-frequency information.
\end{comment}

In this paper, we present a low-latency, lightweight, high-accuracy network for the dense depth completion. Specifically, we propose a novel dual-stream attention module (S\&C enhancer) which contains both spatial-wise and channel-wise attention module as enhancer to boost the representational power of encoder network. The proposed module is based on a new finding that each channel in the features generated by our encoder network responds to different distance range. We argue that different channels should be paid different attention. For example, the encoder network should pay more attention to the channels which are corresponding to close distance if one frame contains only near objects. Thus, it can be excepted that a channel-wise attention operation should benefit the performance. It should be pointed out that recent efforts tried on channel-wise attention module [10, 16, 17] have demonstrated their great boost power in image semantic segmentation, classification and super-resolution tasks. 

Our network and S\&C enhancer are shown in Figure \ref{fig:architecture} and Figure \ref{fig:scmodule}. The spatial-wise attention is utilized to capture long-range dependencies. The channel-wise attention is adopted to reassign the weights of features generated by the encoder network. Moreover, recent works have demonstrated that a local refinement module guided by sparse depth map could refine the generated coarse output significantly \cite{cheng2018depth, qiu2018deeplidar:, van2019sparse}. Inspired by these works, we adopt a simple cascaded hourglass network which takes coarse depth map and sparse depth map as inputs to serve as a local refinement network. Our main contributions can be summarized as threefold:
\begin{itemize}
\item A novel lightweight coarse-to-fine network is proposed for the dense depth completion. Our approach has achieved comparable performance on KITTI benchmark dataset \cite{geiger2013vision} with SOTA approaches in the main metric (RMSE) but outperform it in all other metrics (iRMSE, iMAE, MAE and Runtime) with a $4$ times faster speed. 
\item A spatial-and-channel-wise enhancer is proposed to enhance the representational power of lightweight encoder network. Experimental results demonstrate that the proposed S\&C enhancer can be plugged into other existing depth completion networks easily and boost their performance significantly with negligible additional parameters.
\item It is found that each channel in the features extracted by our encoder network responds to different distances. This new finding may be essential for explaining the working mechanism of deep neural network based depth completion approaches and can help other researchers generate novel ideas to boost the performance.
\end{itemize}

\section{Related Work} 
\label{sec:related_work}
In this section, we summarize existing works on dense depth completion, efficient neural networks, and attention mechanism.
%%%%%%%%%%%%%%%%%%%%%%%%%%%%%%%%%%%%%%%%%%%%%
\subsection{Dense Depth Completion}
Estimating a dense depth map from a single-view color image and its correlated sparse depth map becomes more and more popular in both the robotics and computer vision communities. Early works on depth completion task are mainly optimization-based \cite{park2014high-quality}  or filter-based methods \cite{richardt2012coherent}. Recent approaches have achieved significant improvement on accuracy via deep encoder-decoder neural networks. A fully convolution-based method was proposed to synthesize dense depth map in \cite{uhrig2017sparsity}  which takes an RGB image and its related sparse depth map as inputs. \cite{cheng2018depth} proposed a convolution spatial propagation network and further boosted the performance on KITTI benchmark dataset. Beyond fully supervised approaches, \cite{ma2018self-supervised} proposed a self-supervised training framework that requires only sequences of color and sparse depth images, without the need for dense depth labels. Recently, an impressive work employed a network to estimate surface normal as the intermediate supervisor to produce high-accuracy dense depth map \cite{qiu2018deeplidar:}. However, all of these methods focus heavily on attaining higher accuracy at the cost of increasing complexity. Different from these works, we focus on designing a network to estimate high-precision dense depth map with a light-weight architecture so as to make sure our algorithm can be applied in a real-time system.
%%%%%%%%%%%%%%%%%%%%%%%%%%%%%%%%%%%%%%%%%%%%%%%%
\subsection{Efficient Model Designs}
%%%%%%%%%%%%%%%%%%%%%%%%%%%%%%%%%%%%%%%%%%%%%%%%
There has been significant effort in prior works to design efficient networks. For image classification, the \textsl{MobileNet} \cite{howard2017mobilenets:} can achieve comparable top-5 accuarcy with \textsl{VGG-16} \cite{simonyan2015very} on ImageNet with a $33$ times fewer weight. Then, \textsl{ShuffleNet} \cite{hluchyj1991shuffle} further boosted the performance by adopting channel shuffle operation. By taking advantage of inverted residual structure, \textsl{MobileNetV2} \cite{sandler2018mobilenetv2:} can outperform \textsl{MobileNet} and \textsl{ShuffleNet}  with even fewer parameters. Most recently, \textsl{MobileNetV3} \cite{howard2019searching} further improved the performance of efficient network by introducing the squeeze-and-excitation module. For object detection, \textsl{YOLO-V3} \cite{redmon2018v3:} runs $8$ times faster than \textsl{Faster-RCNN} \cite{ren2017faster} and can perform higher mean average precision. For semantic segmentation, \textsl{ERFNet} \cite{romera2018erfnet:} employed the residual factorized convolution module, and can outperform \textsl{FCN} \cite{long2015fully}  with  $20.8$ times faster running speed.  

Among these tasks, the image segmentation is most similar to the dense depth completion, as both of them can be considered as pixel-to-pixel translation problem, which is mainly solved by the encoder-decoder network.  However, suffering from the low representational power of the light-weight encoder network, a light-weight encoder-decoder network can hardly achieve comparable performance with heavy-weight ones. Therefore, designing a powerful feature enhancing module with few additional parameters for a light-weight encoder network is crucial for a light-weight depth completion network.
%%%%%%%%%%%%%%%%%%%%%%%%%%%%%%%%%%%%%%%%%%%%%%%%
%%%%%%%%%%%%%%%%%%%%%%%%%%%%%%%%%%%%%%%%%%%%%%%%
\subsection{Attention Modules}
Attention mechanisms have been widely applied in many tasks \cite{fu2018dual,kim2018ram:,vaswani2017attention} as they can measure long-range dependencies. Specifically, \cite{vaswani2017attention} firstly proposed the self-attention mechanism for neural language processing, which can capture global dependency and contains only a few trainable parameters.  Meanwhile, \cite{wang2018non-local} introduced the self-attention mechanism into computer vision tasks and redesigned it into non-local operation, which captures spatial-wise interrelationship among all pixels in one channel. Beyond spatial-wise attention mechanisms, many works focused on measuring the channel-wise correlation to boost the representational power of backbone network. For instance, squeeze-and-excitation (SE) module \cite{hu2018squeeze-and-excitation} was proposed to reassign weights of different channels by utilizing the descriptor of global distribution. However, the global descriptor used in this article is generated by the simple average pooling, which may weaken the high-frequency components of images. The CAM \cite{kim2018ram:} pointed out the same issue and proposed a global variance pooling based SE module to improve their performance on the super-resolution.  Recently, \cite{fu2018dual} proposed a novel channel-wise attention module, which is similar to non-local operation to boost the performance on semantic segmentation. However, many large matrix operations in this module lead to overhead computation. In this paper, we modify the SE module to produce channel-wise attention and utilize the modified non-local operation to improve the receptive filed of the encoder network. Moreover, we introduce a new approach to joint spatial-wise attention and channel-wise attention efficiently by adding minimal additional parameters.
%%%%%%%%%%%%%%%%%%%%%%%%%%%%%%%%%%%%%%%%%%%%%%%%
\begin{figure*}[t]
	\centering
	\includegraphics[width=\linewidth]{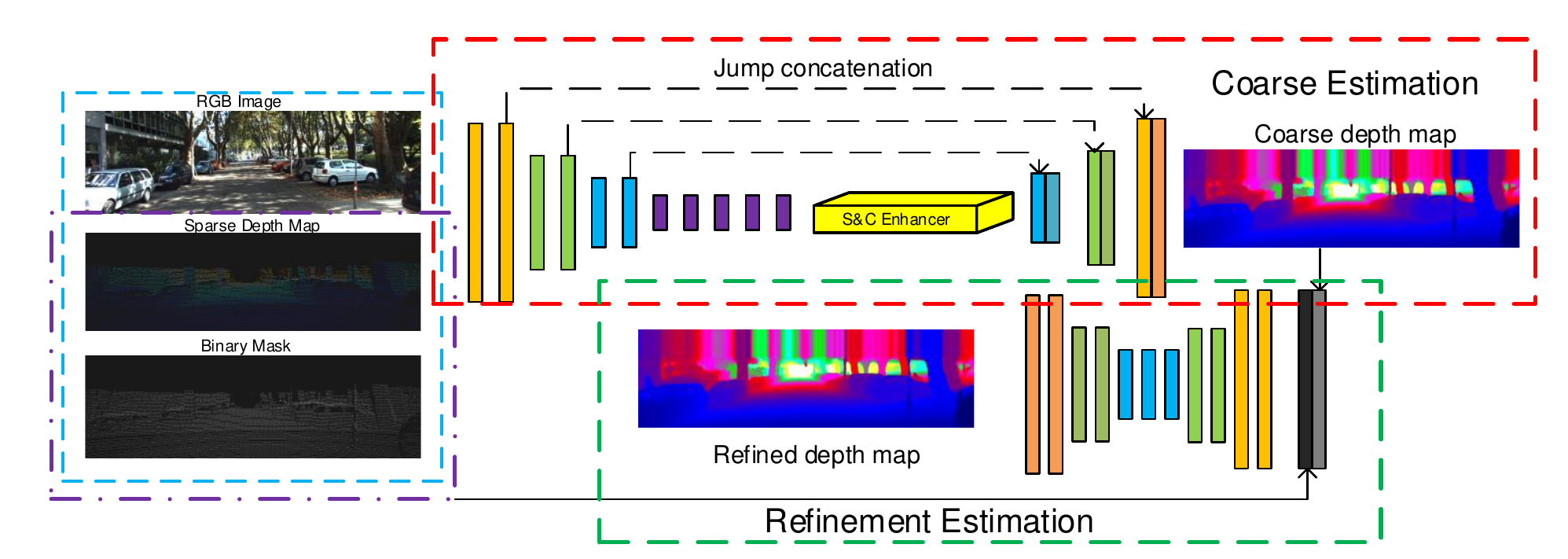}
	\caption{Proposed network architecture. The coarse estimation network is marked with the red box, and the refinement network is marked with the green box. The binary available map is generated from the sparse depth map, if the depth of a pixel is available, the pixel in the binary available map is set to $1$. Otherwise, it is set to $0$.}
	\label{fig:architecture}
\end{figure*}
%%%%%%%%%%%%%%%%%%%%%%%%%%%%%%%%%%%%%%%%%%%%%%%%
\section{Methodology}
\label{sec:methodology}
In this section, we first introduce an S\&C enhancer which can boost the representational power of a light-weight backbone network with minimal additional parameters by fusing a spatial-wise attention module and a channel-wise attention module. Then, we describe our network architecture for the dense depth completion.
%%%%%%%%%%%%%%%%%%%%%%%%%%%%%%%%%%%%%%%%%%%%%%%%
\begin{figure}[t]
	\centering
	\includegraphics[width=\linewidth]{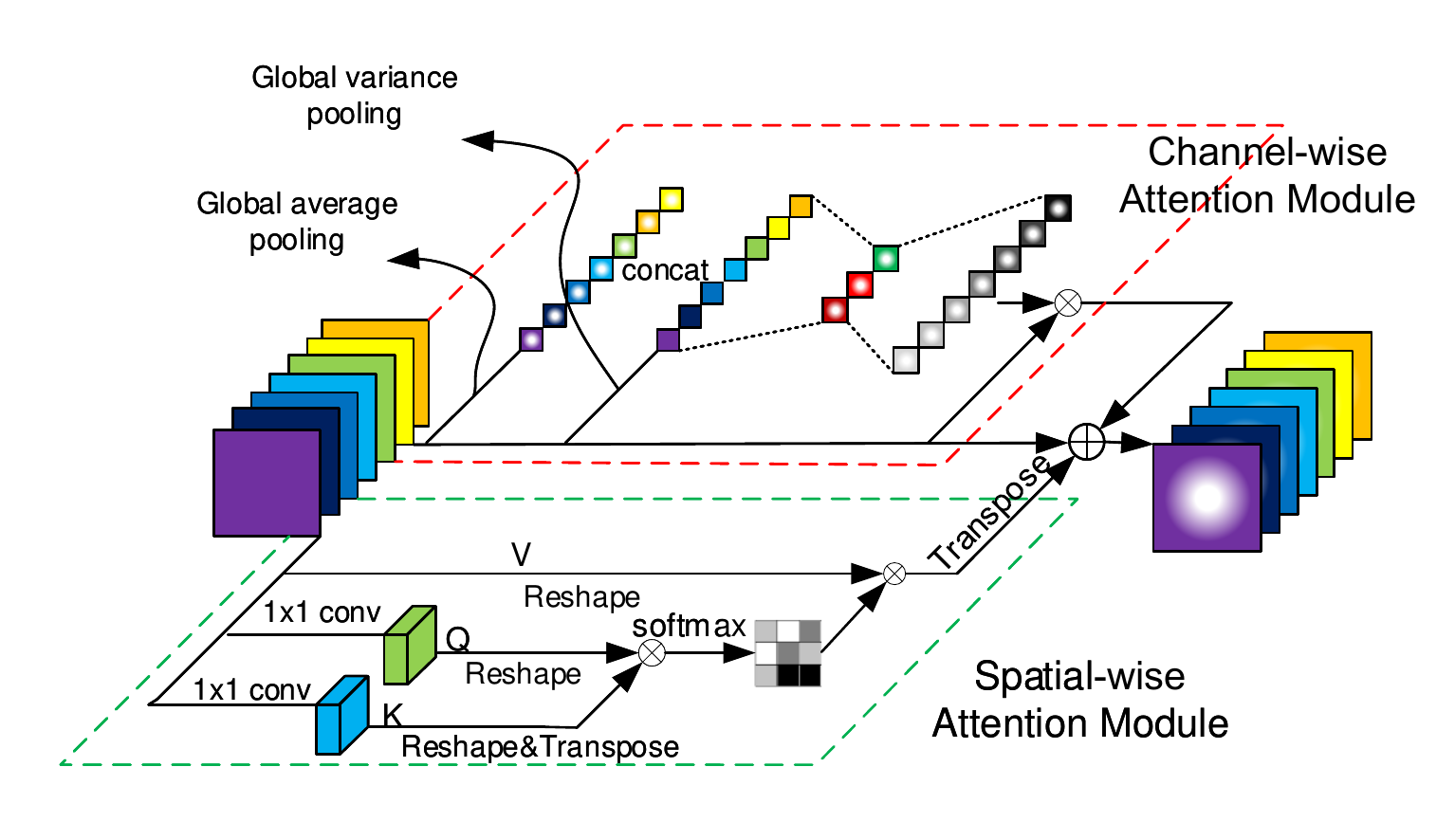}
	\caption{Proposed S\&C enhancer. The spatial-wise enhancer is marked with the green box, and the channel-wise enhancer is marked with the red box.}
	\label{fig:scmodule}
\end{figure}

\subsection{Spatial and Channel Attention Based Enhancer}
\label{SCMODULE}

\subsubsection{Spatial-wise enhancer} \label{subsec:spatial}
The receptive field of our encoder network is limited since we set the output stride of the encoder to $8$ so as to preserve more detail information. As explained in Section \ref{sec:introduction} , the limited receptive field can lead to poor performance. To address this issue, we employee the spatial-wise enhancer, which is similar to non-local operation \cite{wang2018non-local}. Our spatial attention enhancer is illustrated in Figure \ref{fig:scmodule} and is highlighted with the green box.

We denote the feature map as $\mathbf{A} \in \mathbb{R}^{C\times{H}\times{W}}$, where $C$ is the number of feature channels and $H$, $W$ are the height and width of the feature map. For efficiency propose, two $1\times 1$ convolution layers are applied to $\mathbf{A}$ so as to decrease the channel numbers of generated query map (\textbf{Q}) and key map (\textbf{K}). Both \textbf{Q} and \textbf{K} are in shape of $\frac{C}{8}\times{H}\times{W}$. Different from the original non-local operation, our lightweight encoder network generates only 160 channels. Thus it is more sensitive to the information loss caused by channel-compressing operation. This motivates us to employee the identity feature map \textbf{A} as value map \textbf{V} directly. Then, \textbf{Q} and \textbf{K} are reshaped to the size of $\mathbb{R}^{\frac{C}{8}\times N}$, where $N = H\times W$. \textbf{V} is transposed to the shape of $\mathbb{R}^{C\times N}$. Then, the similarity map can be achieved by a matrix multiplication between the transposed \textbf{K} and \textbf{V}. After that, we apply a softmax layer to calculate the spatial attention map $S \in \mathbb{R}^{N \times N}$:
\begin{equation}
s_{ji}=\frac{exp{(K_i^T {Q_j}})}{\sum_{i=1}^{N}exp(K_i^T {Q_j})}
\label{equ.4}
\end{equation}
where $s_{ji}$ measures the impact of the $i_{th}$ position in value map $\mathbf{V}$ on the $j^{th}$ position. Note that the softmax layer indicates that more similar feature representations of the two position contribute to a greater correlation between them. Then we perform a matrix multiplication between $\mathbf{V}$ and the attention map $\mathbf{S}$ to generated the enhanced feature map $\mathbf{E}$:
\begin{equation}
E_j = \sum_{i=1}^{N}(s_{ji}V_i)
\label{equ.5}
\end{equation}
It is worth noting that the non-local operation is different from a fully-connected layer. The non-local operator (2) computes responses based on relationships between different locations, whereas the fully connected layer utilizes learned weights. Finally, $E$ is reshaped back to $\mathbb{R}^{C\times{H}\times{W}}$. It can be inferred from Equation \ref{equ.5} that the spatial enhancer has a global contextual view, and similar represented features achieve mutual gains.

\subsubsection{Channel-wise enhancer.} \label{subsec:channel}
\begin{figure*}[htbp] \centering
	\begin{subfigure}{0.22\linewidth}
		\centering
		\includegraphics[width=1.63in]{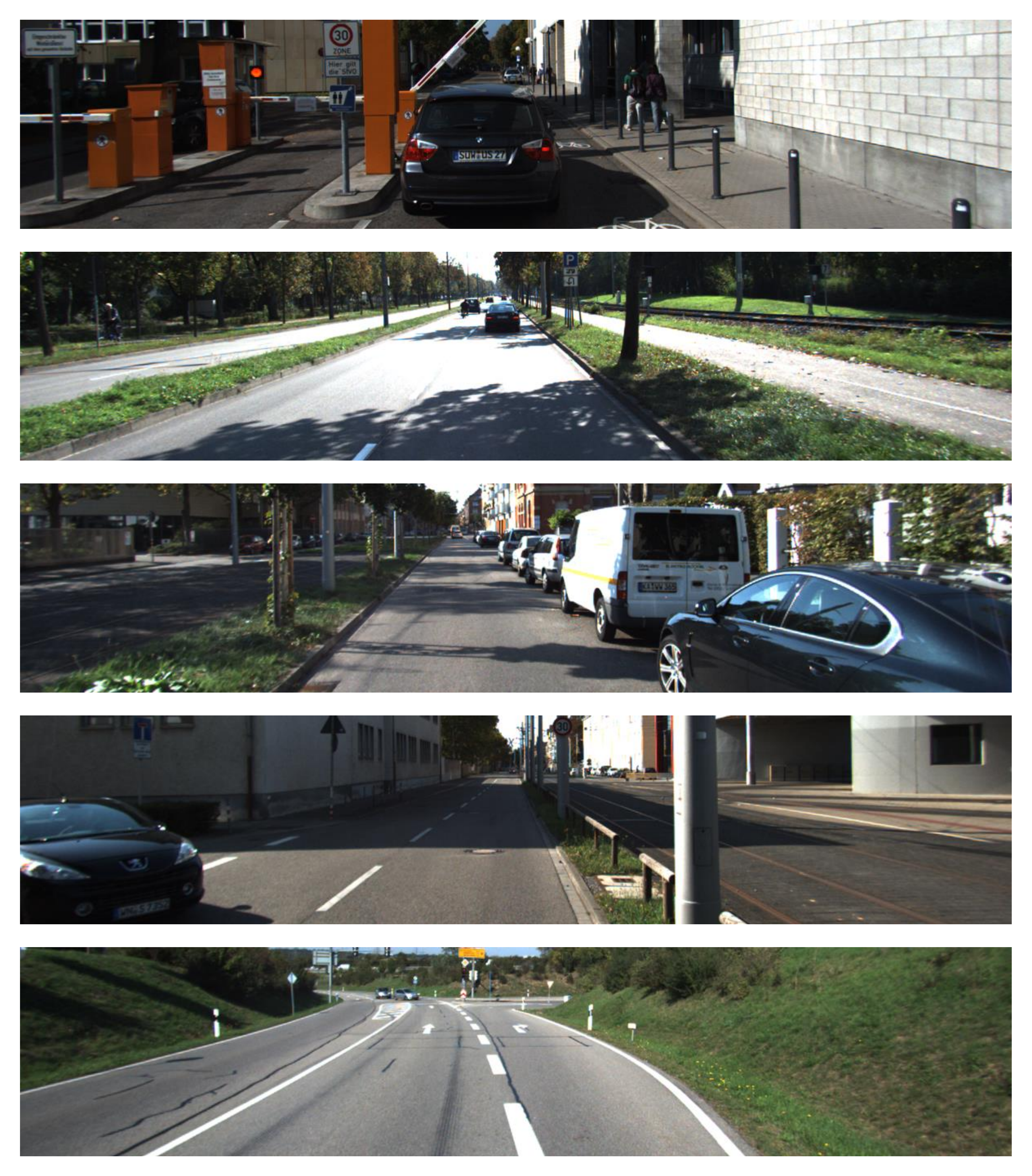}
		\caption{raw rgb image}
		\label{fig:feature-img}
	\end{subfigure} %
	\begin{subfigure}{0.22\linewidth}  
		\centering  
		\includegraphics[width=1.63in]{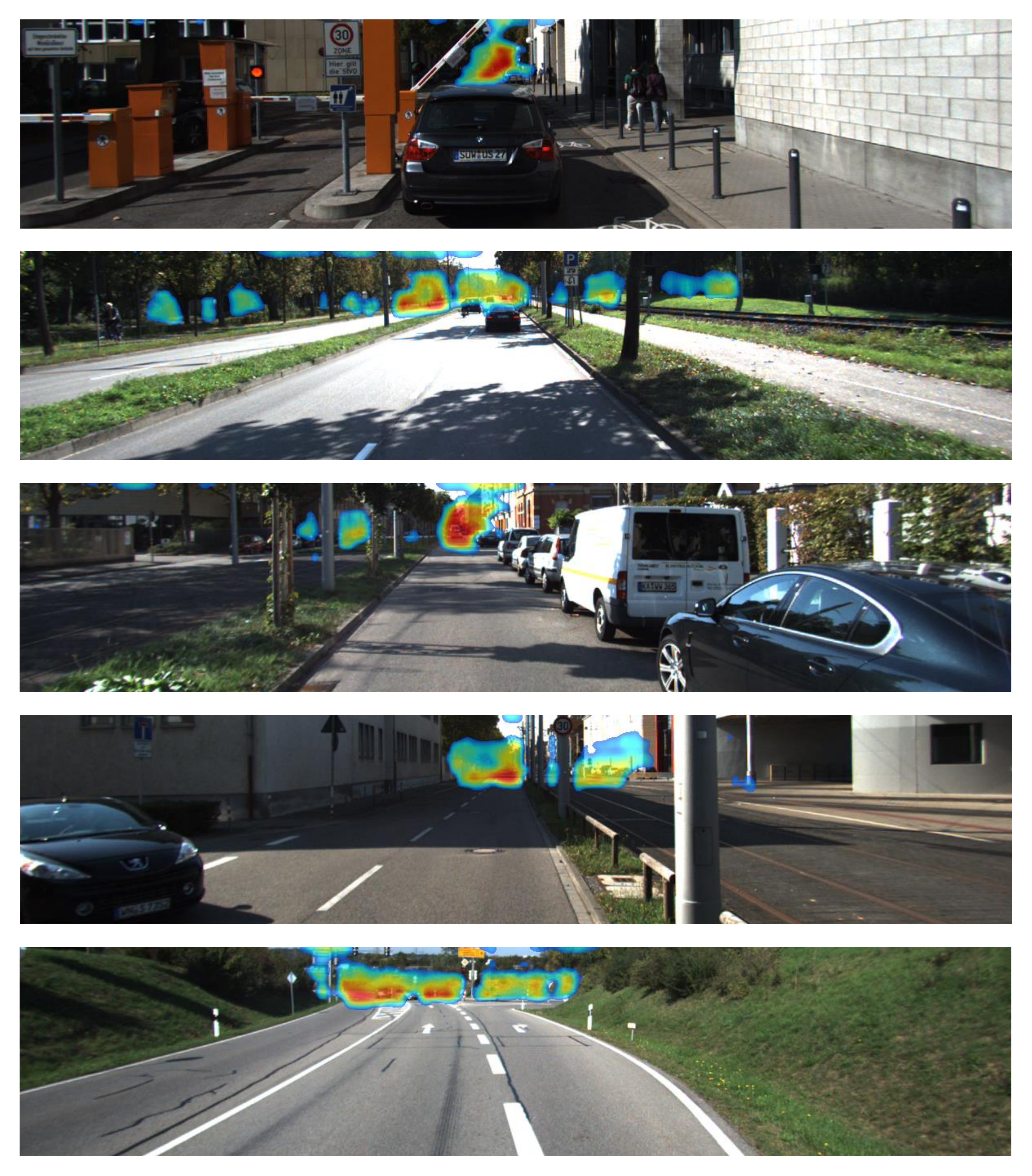}
		\caption{channel 18th}  
		\label{fig:feature-30}
	\end{subfigure} 
	\begin{subfigure}{0.22\linewidth}  
		\centering  
		\includegraphics[width=1.63in]{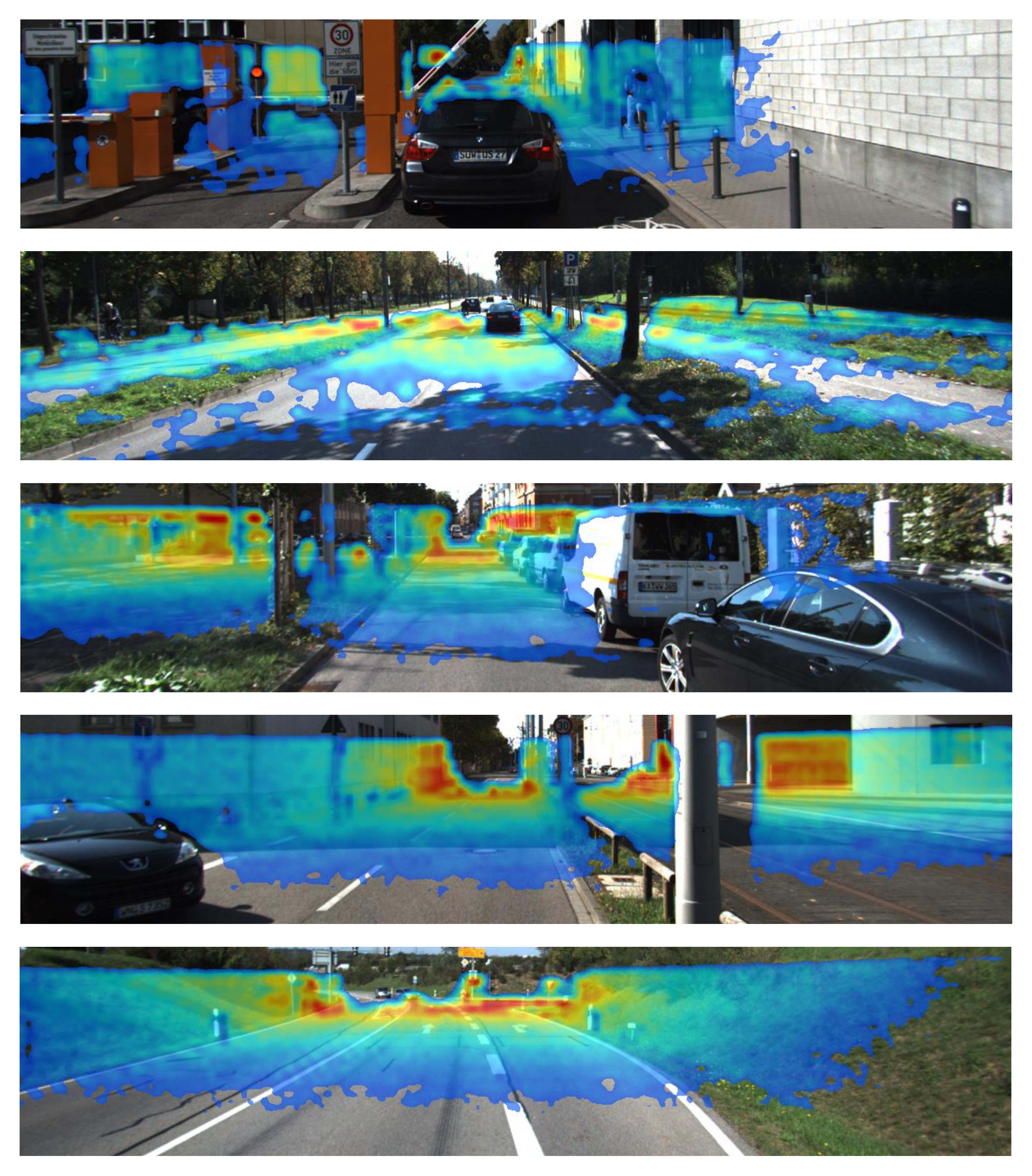}
		\caption{channel 32th}
		\label{fig:feature-160}
	\end{subfigure} 
	%\vspace{0pt} 
	\begin{subfigure}{0.22\linewidth}
		\centering  
		\includegraphics[width=1.63in]{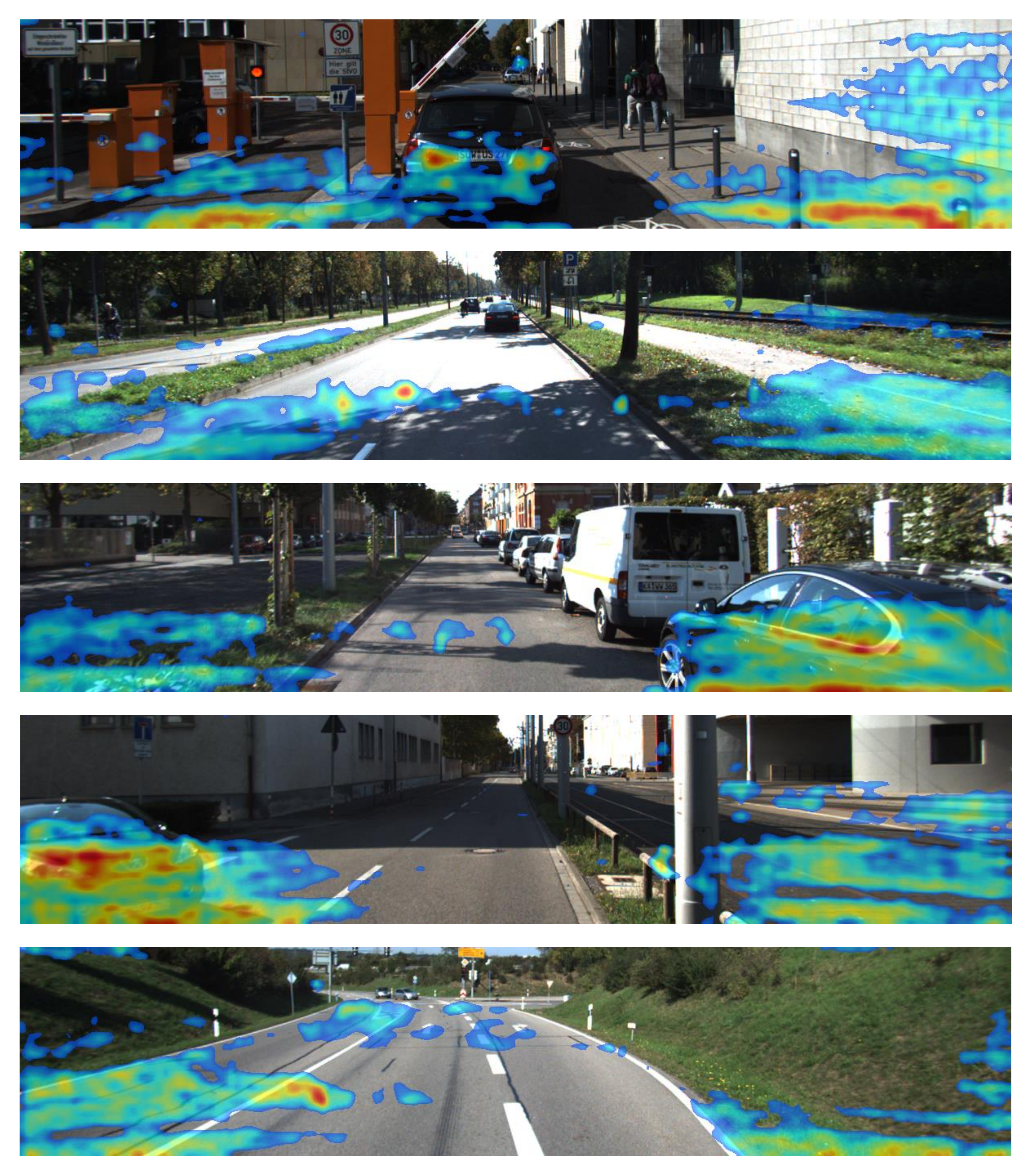}
		\caption{channel 56th}
		\label{fig:feature-110} 
	\end{subfigure}
	%\vspace{0pt} 
	\begin{subfigure}{0.080\linewidth}
		\centering
		\includegraphics[height=2.0in]{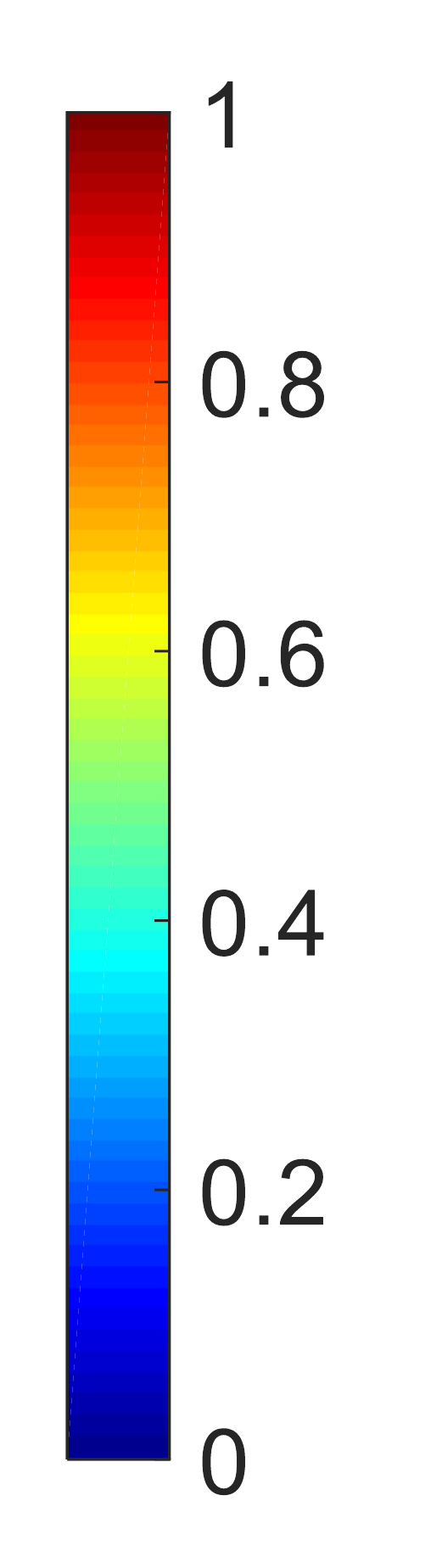}
		%	\caption{channel 56th}
		\label{fig:colorbar} 
	\end{subfigure}
	\caption{Visualization of the intermediate feature map. As shown in the figure above, the 18th channel of our feature map response to long-distance objects, the 32nd feature map response to middle-distance objects and the 56th response to near objects.}
	\label{fig:feature-all}
\end{figure*}

Our study on feature map generated by the encoder network shows that each channel in the feature maps has a different response amplitude to different distance range. Some examples of the activation map are shown in Figure \ref{fig:feature-all}. Intuitively, different channels should be paid different attention. For example, if one frame mainly contains close objects, the channels response to close distance should be paid more attention. This observation motivates us to adopt a modified squeeze-and-extinction (SE) module to reassign the weights of different channels.  

The mean of one channel can be considered as a descriptor of the conten richness in the corresponding distance range. Utilizing the mean of channels helps the SE module producing attention map with the interrelationship of content richness in different channels. Meanwhile, sharp boundaries such as edges are also essential for pix-to-pixel translation problem. Thus high-frequency information should also be considered. For example, if one channel contains complexity boundaries, less attention on it may cause blurry boundaries in the output. The variance of one channel can preserve more sharp boundaries information. Therefore, utilizing both variance and mean of channels instead of only one of them as a global descriptor of SE module is a more robust way in our task. With this consideration, we adopt a modified SE module to serve as our channel-wise enhancer. The detail on our modified SE module is given as follows:

\textbf{Squeeze.} Now we assume that the features need to be enhanced as $\textbf{A} \in \mathbb{R}^{C\times{H}\times{W}}$. To measure the channel interrelationship, we first generate the global channel descriptor by utilizing both of global average pooling and global variance pooling. Thus, the statistic of channel can be written as $\mathbf{z} = [\mathbf{z_c^{mean}}  \; \mathbf{z_c^{var}}] \in \mathbb{R}^{2C}$. Here $\mathbf{z_c^{mean}} \in \mathbb{R}^{C}$ and $\mathbf{z_c^{var}} \in \mathbb{R}^{C}$ are the mean and variance of all channels. The statistic is archived by shrinking $\mathbf{A}$ through its spatial dimensions $H\times W$. Then, the $c_{th}$ element of $\mathbf{z_c^{mean}}$ and $\mathbf{z_c^{var}}$ are calculated by:
\begin{equation}
z_c^{mean} = F_{me}(\textbf{a}_c) = \frac{1}{H \times W} \sum_{i=1}^{H} \sum_{j=1}^{H} a_c(i,j)
\end{equation}
\begin{equation}
z_c^{var} = F_{va}(\textbf{a}_c) = \frac{1}{H \times W} \sum_{i=1}^{H} \sum_{j=1}^{H} (a_c(i,j) - z_c^{mean})^2
\end{equation}
where $\textbf{a}_c$ is the $c_{th}$ channel in the features $\mathbf{A}$ and $a_c(i,j)$ is the feature at position $(i,j)$ in $\textbf{a}_c$.

\textbf{Excitation.} The channel-wise descriptor is utilized to regress the weights of channels with an excitation operation:
\begin{equation}
\textbf{s} = \textbf{F}(\textbf{z,W}) = \sigma(g(\textbf{z,W})) = \sigma(\textbf{W}_2\delta(\textbf{W}_1 \textbf{z}))
\end{equation}
where $\mathbf{s} \in \mathbb{R}^C$ is the generated weights of channels in $\mathbf{A}$. $\textbf{W}_1 \in \mathbb{R}^{\frac{2C}{r} \times 2C}$ and $\textbf{W}_2 \in \mathbb{R}^{C \times \frac{2C}{r}}$ are the parameters of two non-biased fully-connected layers. $\delta$ and $\sigma$ are the ReLU and Sigmoid activation layer. $\mathbf{z}$ is generated by concatenating by the global mean statics $\mathbf{z}^{mean}$ and the global variance statics $\textbf{z}^{var}$. The first FC layer reduce the dimension from $2C$ to $\frac{2C}{r}$ with reduction ratio $r$. Finally, the channel wise enhanced features $\textbf{X} \in \mathbb{R}^{C\times{H}\times{W}}$ is obtained by resealing $\textbf{A}$ with the weights $\mathbf{s}$:
\begin{equation}
\textbf{x}_c = F(\textbf{a}_c,s_c) = s_c \textbf{a}_c
\end{equation}
where $\textbf{x}_c$ is the $c_{th}$ channel in the channel-wise enhanced feature $\textbf{X}$.

\subsection{Fusing}
The proposed channel-wise enhancer and spatial-wise enhancer utilize information from inter-channel and intra-channel relationship. It is crucial  to fuse the enhanced feature map properly. Directly concatenating the two enhanced feature maps can exploit the benefits of both mechanisms mostly but could introduce too many additional parameters. To make use of both mechanisms simultaneously, we further multiply the two enhanced features by trainable scale parameters and add back the identity feature map. Therefore, the final enhanced feature map is given by:
\begin{equation}
\textbf{Y} = \lambda \textbf{E} + \gamma \textbf{X} + \textbf{A}
\end{equation}
where the $\textbf{Y}$ is the final output of S\&C enhancer, $\textbf{E}$ is the output of the spatial-wise attention module, $\textbf{X}$ is the output of channel-wise attention module and the $\textbf{A}$ is the identity feature map. Beyond the proposed fusing strategy, we also try other strategies to fuse the features, and the controlled experiment is shown in Section \ref{sec:ablation}.

%%%%%%%%%%%%%%%%%%%%%%%%%%%%%%%%%%%%%%%%%%%%%%%%
\subsection{Coarse-to-Fine Network Architecture}
Our proposed coarse-to-fine framework is illustrated in Figure \ref{fig:architecture}. The coarse estimation network is shown in the red box, and the refinement network is shown in the green box. In the coarse estimation network, the MobileNetV3-Large \cite{howard2019searching} is employed as an encoder to extract the features from the input (raw RGB image,  sparse depth map, and available binary mask). The output stride of the encoder is set to $8$. Then, our proposed spatial and channel (S\&C) enhancer, which has been illustrated in section \ref{SCMODULE}, is plugged after encoder to enhance the extracted features. Then, the enhanced features are feed into the decoder, which consists of three up-projection units as introduced in \cite{romera2018erfnet:} to gradually increase the feature resolution. To improve the detail performance of the network, we concatenate the low-level features from encoder with the up-projected features at the same resolution. It should be pointed out that we do not concatenate the low-level features from encoder directly because the corresponding low-level features usually contain a large number of channels. Instead, we apply other $1\times 1$ convolutions on the low-level features to reduce the number of channels, and then concatenate it with the corresponding decoder features.

We adopt a simple casted hourglass module as a refinement network to improve the performance. Then, the coarse estimation output, the sparse depth map and the binary available mask is concatenated as the input of the refinement network. Note that the binary available mask also provides a significant amount of knowledge, it can help the network know which missing areas it should focus on. However, the RGB image is not fed into the refinement network as we expect the refinement network to focus on the available pixels in the sparse depth map. With the guide of the sparse depth map, the refinement network serve as a local patch which can refine the coarse generated dense depth map with the guidance of high-confidence sparse depth map.
%%%%%%%%%%%%%%%%%%%%%%%%%%%%%%%%%%%%%%%%%%%%%%%%
%%%%%%%%%%%%%%%%%%%%%%%%%%%%%%%%%%%%%%%%%%%%%%%%
\section{Experiments} 
\label{sec:experiments}
In this section, extensive experiments are conducted to demonstrate the performance of our approach and the boosting power of our proposed S\&C enhancer. We first provide an ablation study of our design to analyze the influence of each component. Then we present an evaluation against existing published work on KITTI on-line benchmark. Finally, we plug S\&C enhancer into the existing network and compare the evaluation result against the original network to demonstrate its boost power.
%%%%%%%%%%%%%%%%%%%%%%%%%%%%%%%%%%%%%%%%%%%%%%%%
\subsection{Datasets and Metrics}
\label{sec:exp-setup}
We train our networks and evaluate their accuracy on the KITTI dense depth completion dataset \cite{geiger2013vision} with the official train/validation data split. The datasets are collected using two HD camera and Velodyne 64-line LiDARs. This dataset includes semi-dense annotations, RGB image, and sparse depth map. In detail, it contains 86k images for training, 3k images for testing and 4k images for validation. Moreover, all splits are ensured a similar distribution over KITTI scene categories. To evaluate the performance, we utilize the metrics provided by KITTI on-line benchmark. They are:
\begin{itemize}
    \item \textbf{iRMSE}: Root mean squared error of the inverse depth $[1/km]$.
    \item \textbf{iMAE}: Mean absolute error if the inverse depth $[1/km]$.
    \item \textbf{RMSE}: Root mean squared error $[mm]$.
    \item \textbf{MAE}: Mean absolute error $[mm]$.
    \item \textbf{Rumtime}: Time cost for inference one frame $[ms]$.
\end{itemize}

\subsection{Implementation Details}
We implement our model using \textsl{PyTorch} \cite{paszke2017pytorch}. Our framework contains two sub-networks: the coarse estimation network and the refinement network. Both of the two networks are in the encoder-decoder form. In the coarse generation network, we adopt the MobileNetV3-Large \cite{howard2019searching} implementation as our backbone encoder network. It is important to emphasize  that we set the output stride of MobileNetV3-Large to 8 instead of 32. We achieve this by setting the output stride of the $2nd$ and $13th$ \textsl{"bneck"} module to $1$. The backbone network is pre-trained on the ImageNet dataset \cite{deng2009imagenet}. Then, we plugged the proposed S\&C enhancer after the encoder network to enhance the features. After this, we utilize the up projection module proposed in \cite{romera2018erfnet:} to generate the coarse estimation depth map. In the end, we adopt two cascaded stacked hourglass network \cite{newell2016stacked} as the refinement network.

The network is trained with 32-bit floating-point precision. For training, a batch size of 16 and a learning rate of $0.001$ are used. The Adam \cite{kingma2014adam} is employed as optimizer. Furthermore, we crop the inputs to a resolution of $256 \times 1216$ since the LiDAR frame does not provide any information at the top. And the $MSE$ (mean squared error) is employed as the loss function of our model. Then, the loss function is given by:
\begin{equation}
\begin{split}
Loss_{depth}&(pred_{coarse},pred_{refine},d)\\
&=\alpha || \mathbb{I}_{d>0} \cdot{(pred_{coarse} -d)}||_2^2\\ 
&+\beta || \mathbb{I}_{d>0}\cdot{(pred_{refine} -d)}||_2^2
\end{split}
\end{equation}
where $pred_{coarse}$ is the dense depth map generated by the coarse estimation network and $pred_{refine}$ is the refined depth map. Note that we only calculate the pixels whose depth information is provided in the groudtruth semi-dense annotations depth map. The $\alpha$ and $\beta$ are weights of the coarse result and the refined result, which is set to be $0.3$ and $0.7$ respectively.
%%%%%%%%%%%%%%%%%%%%%%%%%%%%%%%%%%%%%%%%%%%%%%%%%%%%%%%%%%
\begin{figure*}[htbp] \centering
	\begin{subfigure}{0.28\linewidth}
		\centering
		\includegraphics[width=2.07in]{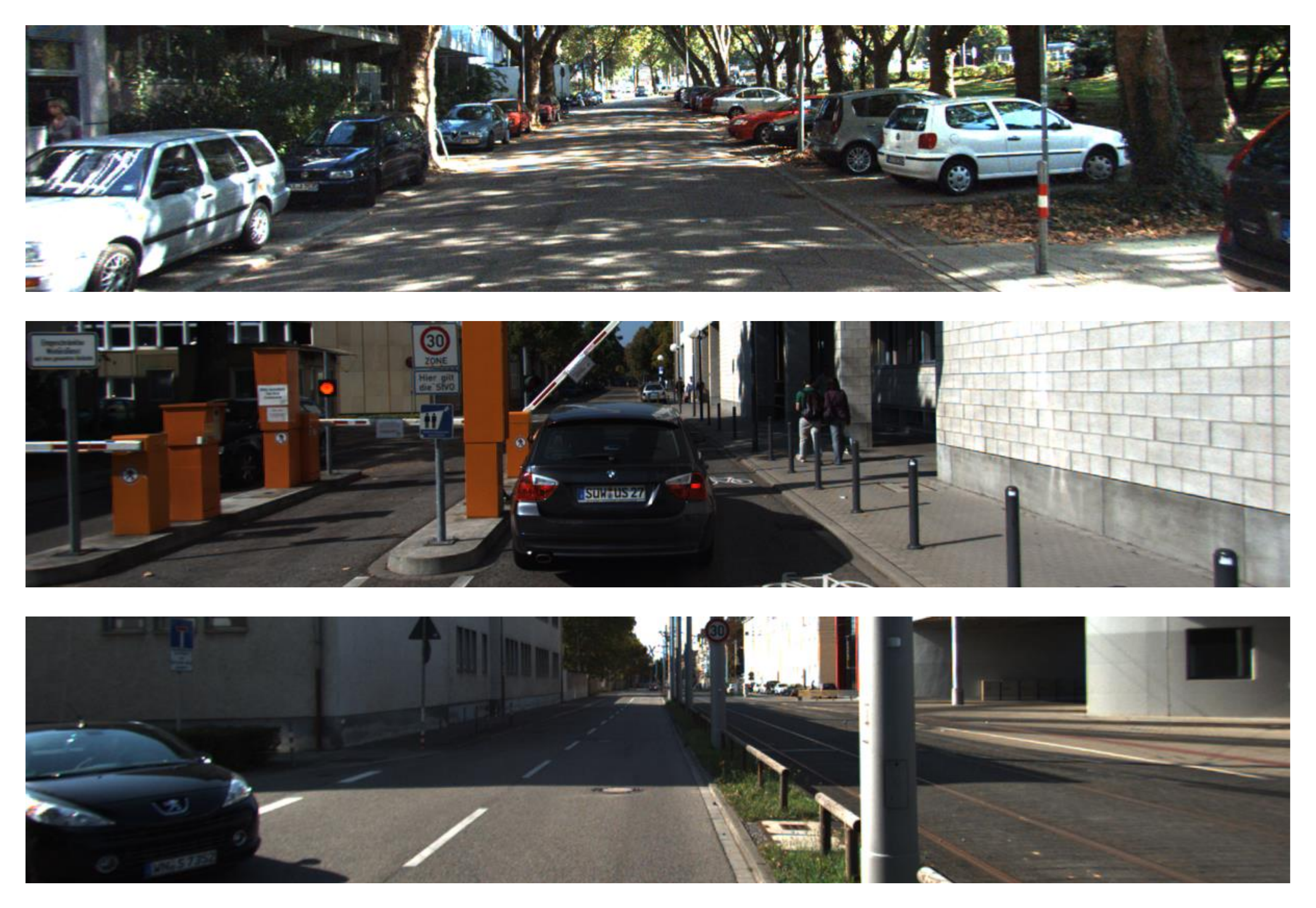}
		\caption{raw RGB image}
		\label{fig:spatial-img}
	\end{subfigure} %
	\begin{subfigure}{0.28\linewidth}  
		\centering  
		\includegraphics[width=2.07in]{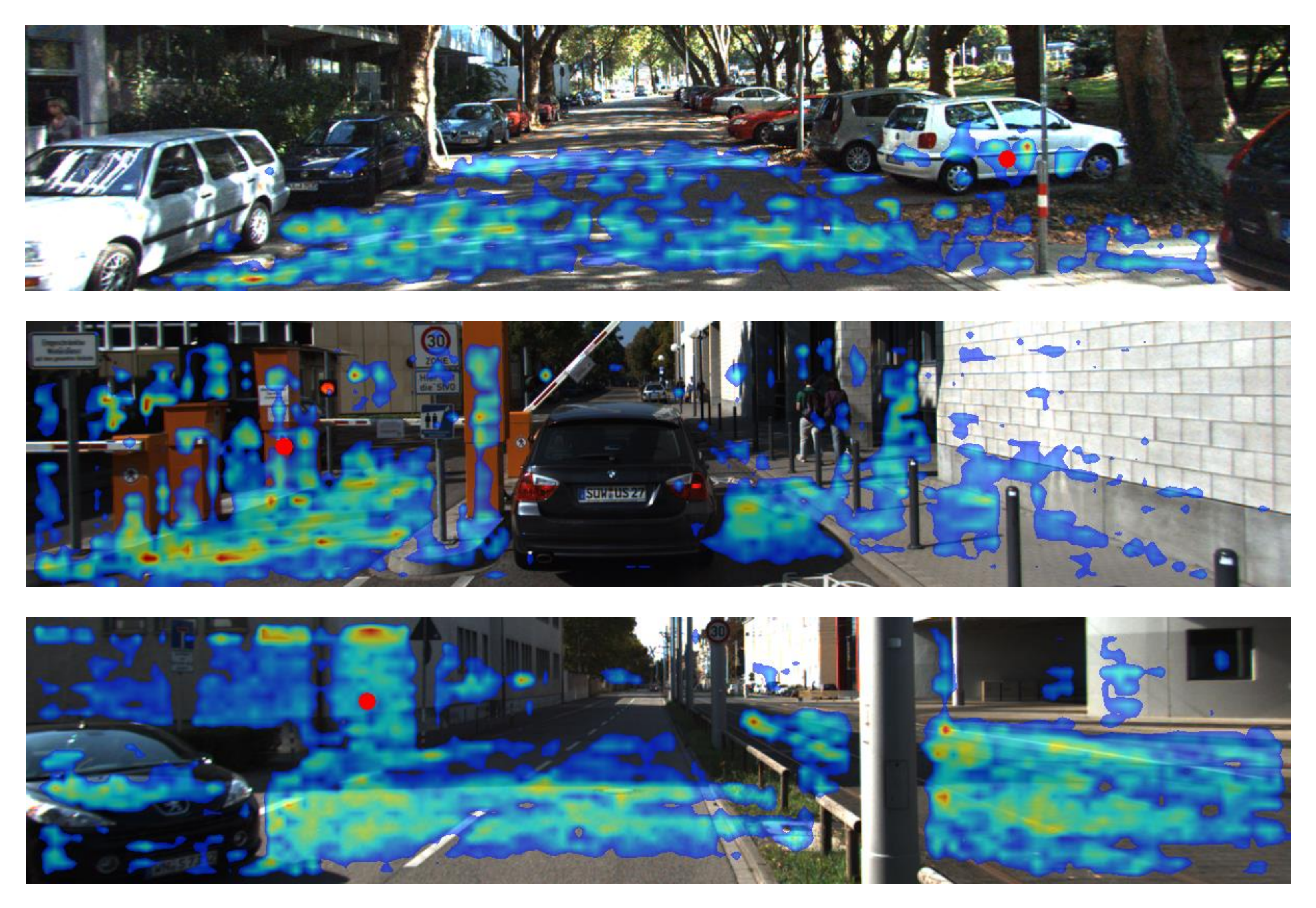}
		\caption{point A}  
		\label{fig:spatial-1}
	\end{subfigure} 
	\begin{subfigure}{0.28\linewidth}  
		\centering  
		\includegraphics[width=2.07in]{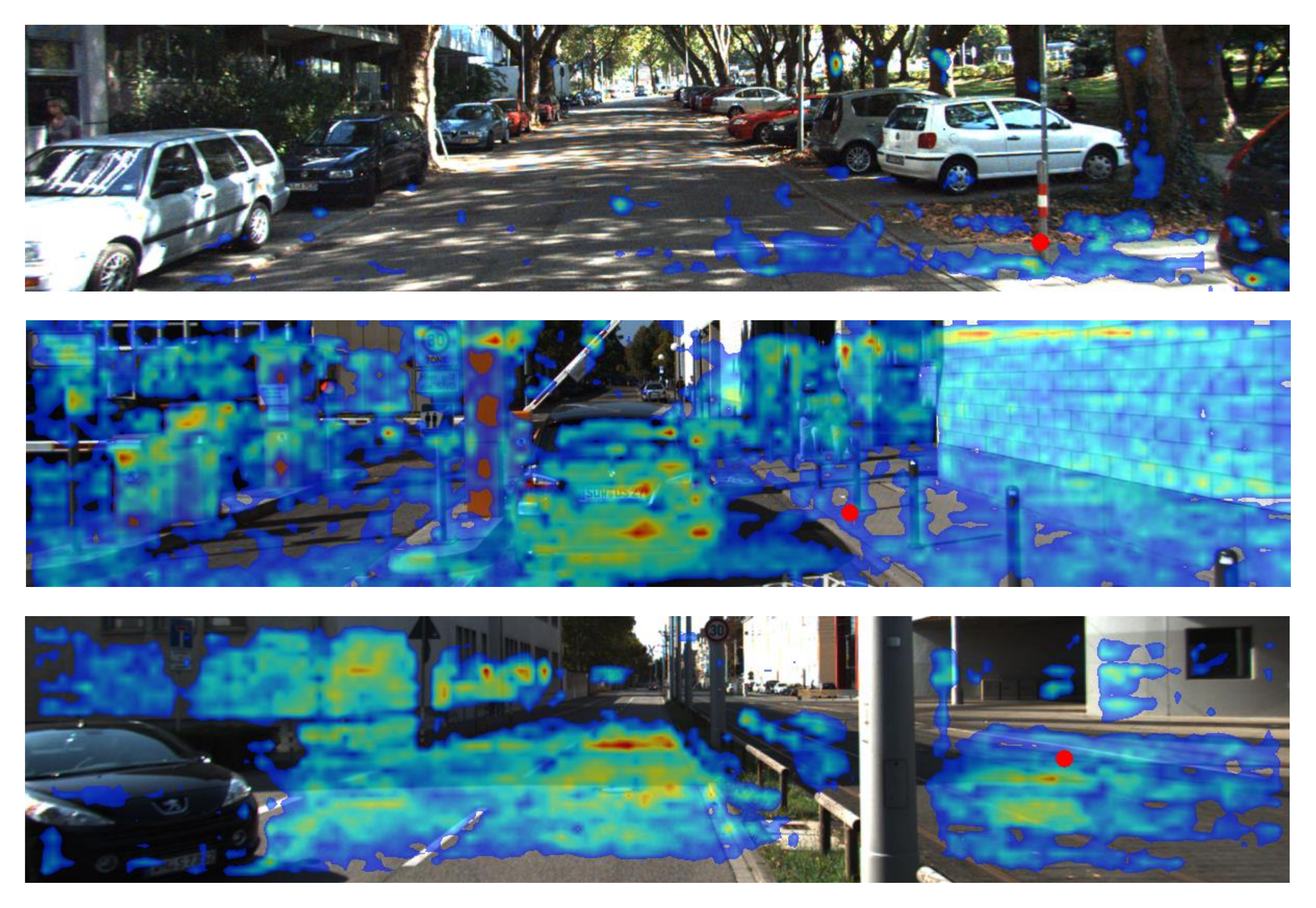}
		\caption{point B}
		\label{fig:spatial-2}
	\end{subfigure}
	\begin{subfigure}{0.06\linewidth}
		\centering
		\includegraphics[height=1.5in]{pics/KITTI_pic/colorbar}
		%\caption{}
		\label{fig:colorbar_position} 
	\end{subfigure}

	\caption{The attention map is visualized by jet colormap and the generated colormap is applied on the RGB image. The corresponding points of visualized attention map are marked with red dots.}
	\label{fig:spatial}
\end{figure*}
%%%%%%%%%%%%%%%%%%%%%%%%%%%%%%%%%%%%%%%%%%%%%%%%
%%%%%%%%%%%%%%%%%%%%%%%%%%%%%%%%%%%%%%%%%%%%%%%%%%%%%%%%%
\begin{table*}[htbp]
	\caption{Ablation study of the our network. The sign \checkmark indicates that this module is added into the framework. The best and the second best records are marked as \textbf{bond} and \color{blue}blue \color{black}.}
	\centering
	% \small
	% \footnotesize
	\scriptsize
	\setlength\tabcolsep{12pt} % default value: 6pt
	\label{tab:ablation}
	\begin{threeparttable}[t]
		%\newcolumntype{M}{>{$\vcenter\bgroup\hbox\bgroup}c<{\egroup\egroup$}}
		\begin{tabular}{| c  *{5}{ c } || r |  }
			\hline
			\textbf{output stride (OS)} & \textbf{skip connection} & \textbf{spatial attention} & \textbf{channel attention} & \textbf{fusing} & \textbf{refinement network} & \textbf{RMSE} \\
			\hline \hline
			$OS = 8$ & \checkmark & \checkmark & proposed & proposed & \checkmark & \color{blue}{786.32}\\
			
			$OS = 8$ & \checkmark & \checkmark & mean & proposed & \checkmark & 793.82\\ 
			
			$OS = 8$ & \checkmark & \checkmark & variance & proposed & \checkmark & 794.40\\
			
			$OS = 8$ & \checkmark & \checkmark & proposed & concat & \checkmark & \textbf{785.25}\\
			
			$OS = 8$ &  & \checkmark & proposed & proposed & \checkmark & 796.13\\ 
			
			$OS = 8$ &  &  & proposed & proposed & \checkmark & 815.32\\ 
			
			$OS = 8$ &  &  &  &  & \checkmark & 831.02 \\ 
			
			$OS = 8$ &  &  &  &  & & 879.33 \\
			\hline \hline
			$OS = 16$ & \checkmark & \checkmark & proposed & proposed & \checkmark & \textbf{794.80}\\
			
			$OS = 16$ & \checkmark & \checkmark & mean & proposed & \checkmark & 801.08\\ 
			
			$OS = 16$ & \checkmark & \checkmark & variance & proposed & \checkmark & 804.79\\
			
			$OS = 16$ & \checkmark & \checkmark & proposed & concat & \checkmark & \color{blue}{795.21}\\
			
			$OS = 16$ &  & \checkmark & proposed & proposed & \checkmark & 817.02\\ 
			
			$OS = 16$ &  &  & proposed & proposed & \checkmark & 821.32\\ 
			
			$OS = 16$ &  &  &  &  & \checkmark & 848.32 \\ 
			
			$OS = 16$ &  &  &  &  & & 884.77 \\
			\hline
			
		\end{tabular}
	\end{threeparttable}
\end{table*}
\subsection{Ablation Studies}
\label{sec:ablation}
To analyze the influence of network components on performance, we conduct a detailed ablation by testing each component in the framework and their combinations to see how they contribute to the estimation precision. The ablation study is done one the KITTI selected cropped validation dataset. The detailed results under different experimental conditions are listed in Table \ref{tab:ablation} and the experiment condition is listed in the first row. The illustration of the experiment condition is listed below:
\begin{itemize}
	\item \textbf{output stride}: this column stands for the output stride of encoder network in coarse estimation network. $OS=8$ indicate we set the output stride of MobileNetV3-Large by setting the output stride of both $2nd$ and $13th$ "\textsl{bneck}" block to $1$. $OS=16$ indicate we set the output stride of MobileNetV3-Large by setting only the $13th$ "\textsl{bneck}" block to $1$.
	\item \textbf{skip connection}: \checkmark indicates that we add skip connection between encoder and decoder in the coarse generation module. On the contrary, Blank indicates the skip connection is not employed.
	\item \textbf{spatial attention}: \checkmark indicates that we plugged the proposed channel attention at the end of MobileNetV3-Large, and blank indicates the spatial attention enhancer is not adopted.
	\item \textbf{channel attention}: "\textsl{proposed}" indicate we plugged the proposed channel-wise attention module into the coarse generation network. "\textsl{mean}" indicate the SE module proposed in \cite{hu2018squeeze-and-excitation} is adopted. "\textsl{variance}" indicate the SE module proposed in \cite{kim2018ram:} is employed. Blank indicates there is no channel attention module in our network.
	\item \textbf{fusing} "\textsl{proposed}" indicate we adopt the proposed strategy to fuse the enhanced features generated by channel and spatial enhancer. "\textsl{concat}" indicate we simply concatenate the spatial and channel enhanced features.
	\item \textbf{refinement network} this column shows we adopt the refinement network or not. \checkmark  indicates that the refinement network is adopted and blank indicate not.
\end{itemize}

By analyzing Table \ref{tab:ablation}, we can conclude that the most effective components in improving final accuracy include using proposed S\&C enhancer and adding the cascade refinement network. Besides, adding skip connections between the encoder and decoder networks in coarse estimation network, decreasing the output stride of backbone also result in substantial improvement. Meanwhile, we test different strategies to fuse the enhanced feature map generated by the spatial enhancer and channel enhancer. The experiment result shows that the concatenation strategy only outperforms our proposed fusing strategy a bit little but introduces much more parameters and a lot of computational cost. Besides, our proposed channel-wise attention mechanism performs better in our task than the SE module proposed in \cite{kim2018ram:} and \cite{hu2018squeeze-and-excitation}, the result can be inferred in lines 1st, 2nd, 3th in the block \textbf{OS = 8} and block \textbf{OS = 16}. At the condition \textbf{OS = 8},  the mean based SE module and variance-based module performs quite similar. However, when \textbf{OS = 16}, the mean based SE module performs much better than the variance-based module. This phenomenon may be caused by the loss of detail information when the \textbf{OS} goes high. Besides, by analyzing the first block and second block in Table \ref{tab:ablation}, we can conclude that a lower \textbf{OS} can also improve the performance in our task

We also tested the additional regularization, such as weight decay, leads to degraded performance. However, adding dropout do improve the performance on validation and testing dataset. Moreover, it is worth noting that alternative decoding method of the enhanced feature map (such as nearest-neighbor interpolation or the bilinear interpolation) does not improve the estimation accuracy.
\begin{figure*}[t] \centering
	\begin{subfigure}{0.32\linewidth}
		\centering
		\includegraphics[width=2.35in]{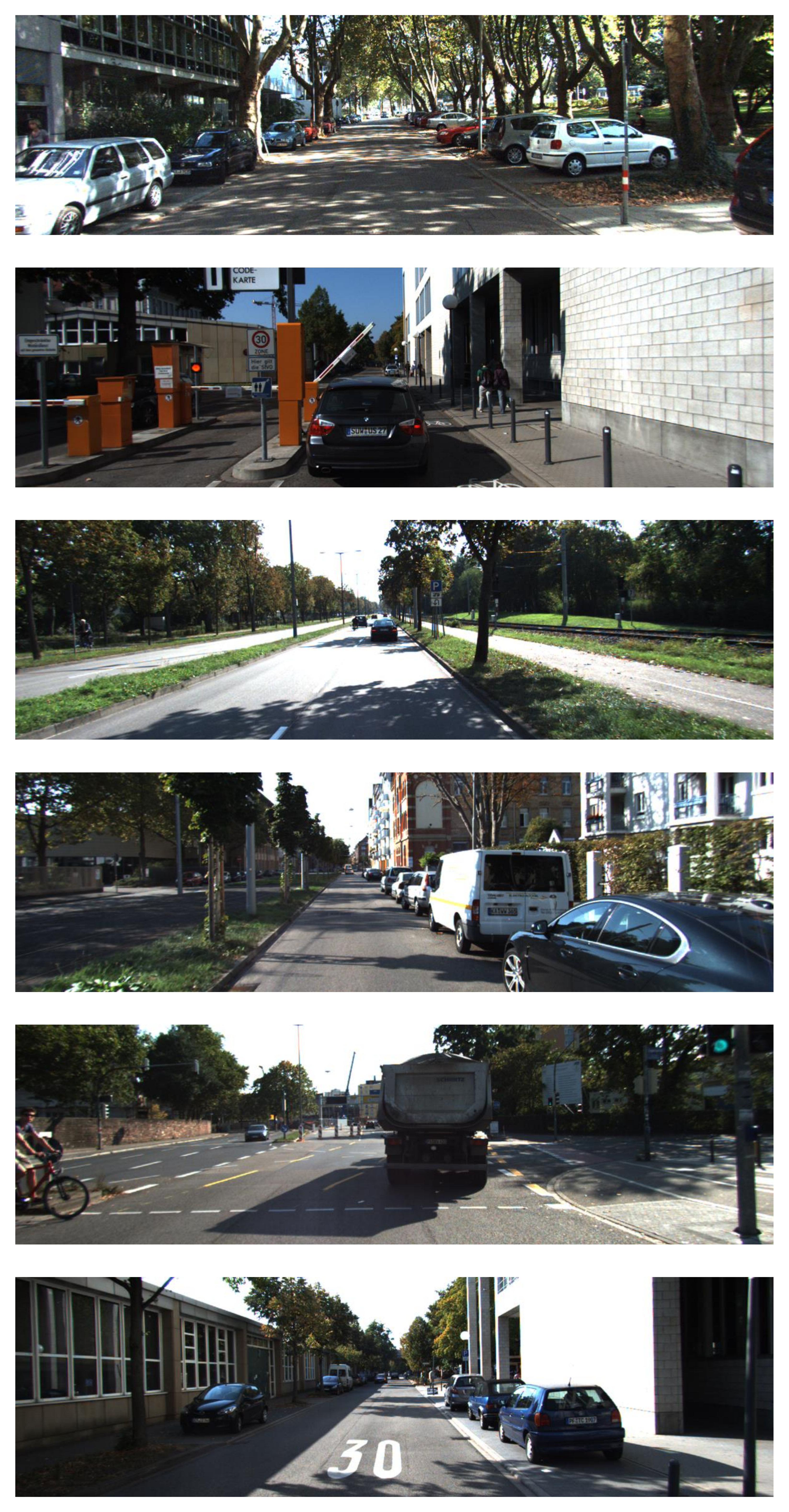}
		\caption{RGB image}
		\label{fig:compare-img-enhance}
	\end{subfigure} %
	\begin{subfigure}{0.32\linewidth}  
		\centering  
		\includegraphics[width=2.35in]{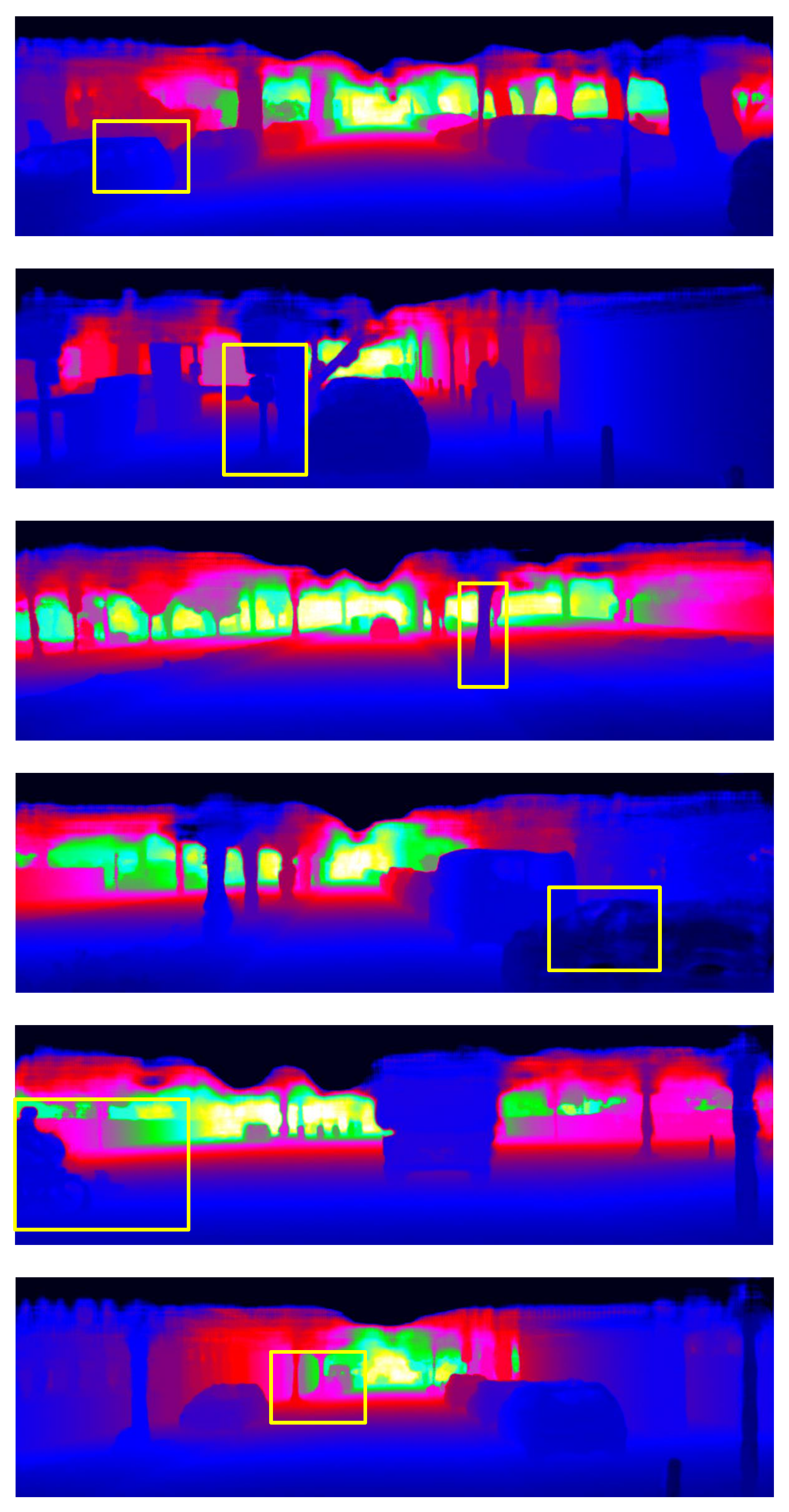}
		\caption{sparse-to-dense}  
		\label{fig:sparse-dense-ori}
	\end{subfigure} 
	\begin{subfigure}{0.33\linewidth}  
		\centering  
		\includegraphics[width=2.35in]{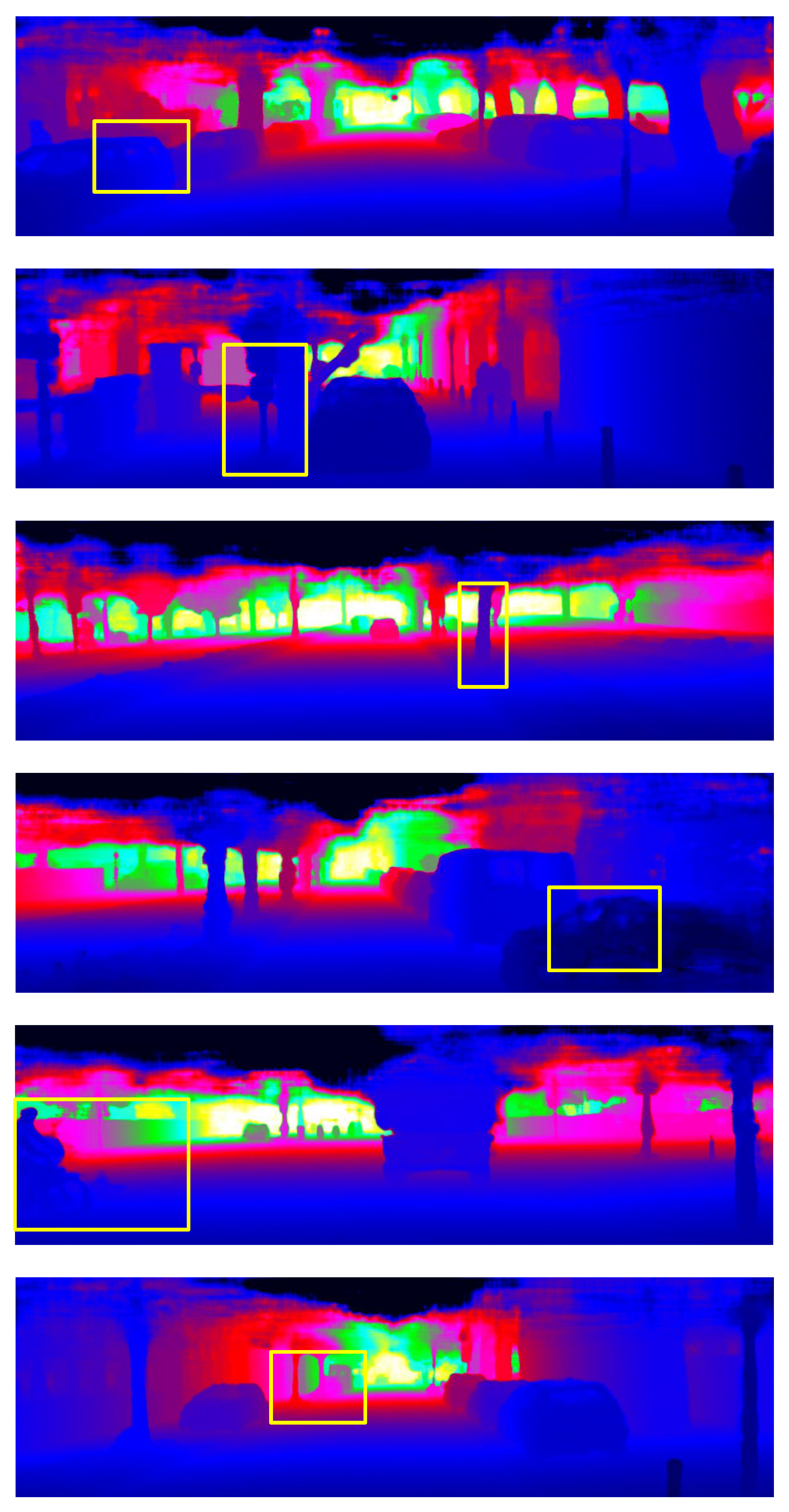}
		\caption{enhanced sparse-to-dense}
		\label{fig:sparse-dense-new}
	\end{subfigure} 
	\caption{Comparison against original sparse-to-dense. The enhanced approach perform not only higher accuracy, but also cleaner boundaries.}
	\label{fig:compare-new}
\end{figure*}
\begin{figure*}[t] \centering
	\captionsetup[subfigure]{labelformat=empty}
	\begin{subfigure}{0.24\linewidth}
		\centering
		\includegraphics[width=1.73in]{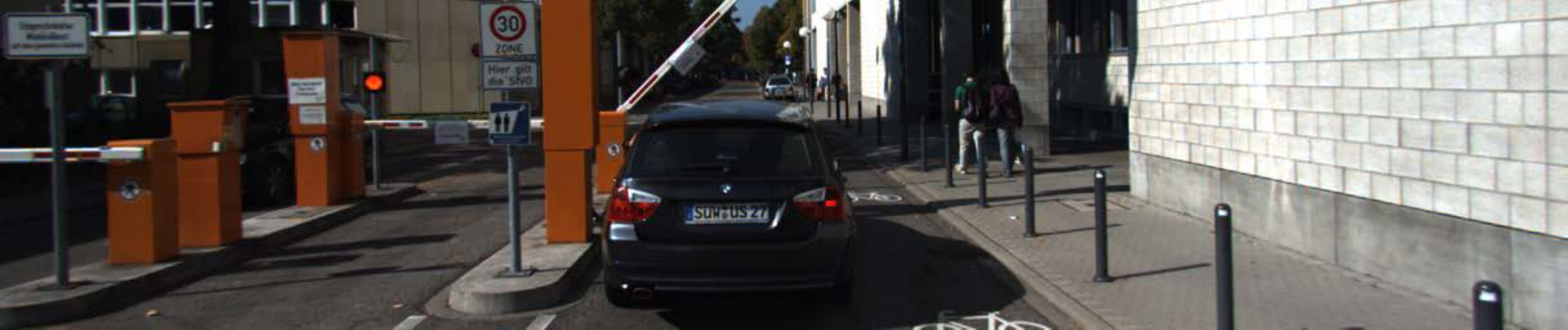}
		\caption{RGB image}
		\label{fig:channel-11}
	\end{subfigure} %
	\begin{subfigure}{0.24\linewidth}  
		\centering  
		\includegraphics[width=1.73in]{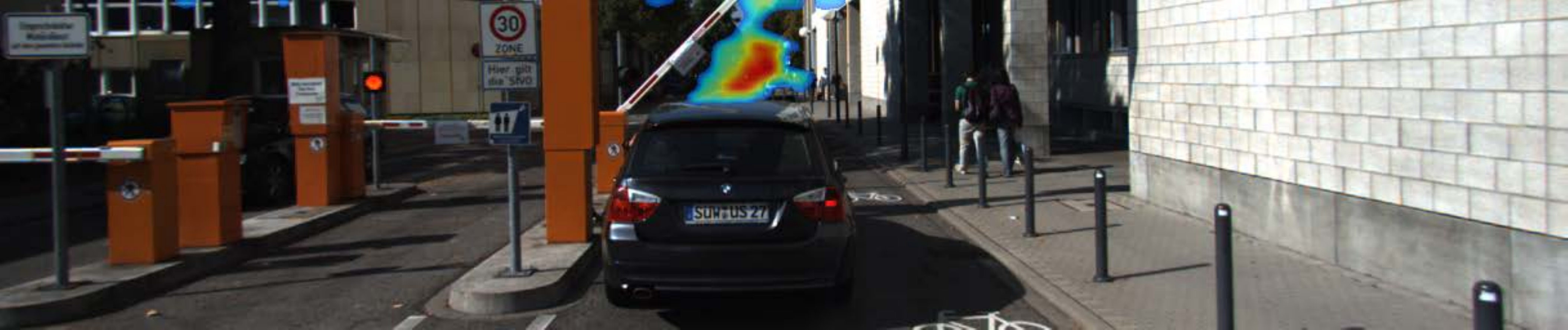}
		\caption{18th channel,$weight=0.879$}  
		\label{fig:channel-12}
	\end{subfigure} 
	\begin{subfigure}{0.24\linewidth}  
		\centering  
		\includegraphics[width=1.73in]{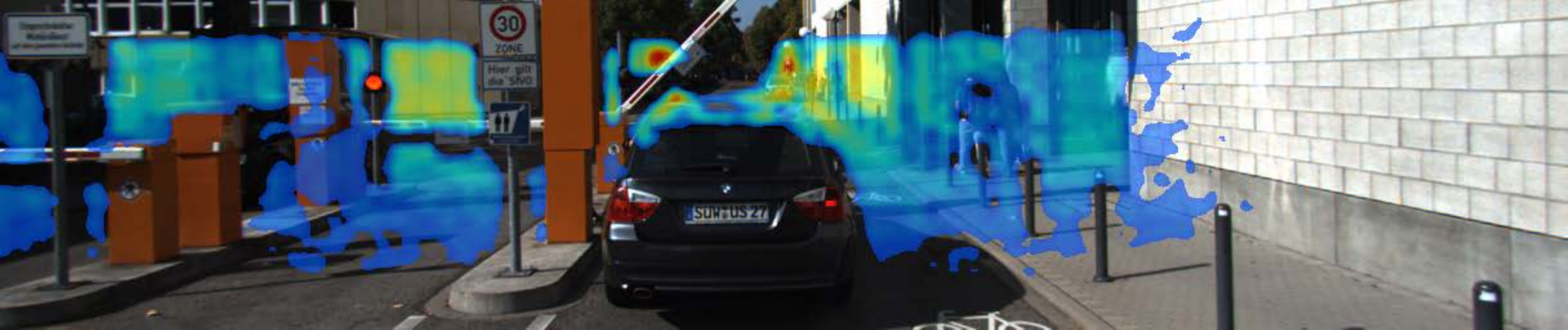}
		\caption{32th channel,$weight=0.999$}  
		\label{fig:channel-13}
	\end{subfigure}
	\begin{subfigure}{0.24\linewidth}  
		\centering  
		\includegraphics[width=1.73in]{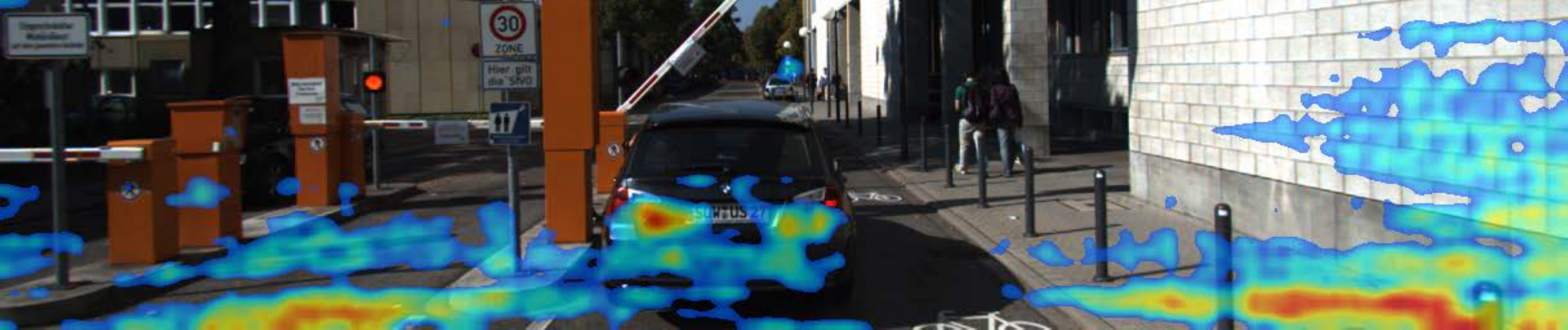}
		\caption{56th channel,$weight=0.569$}  
		\label{fig:channel-14}
	\end{subfigure}
	
	\begin{subfigure}{0.24\linewidth}
		\centering
		\includegraphics[width=1.73in]{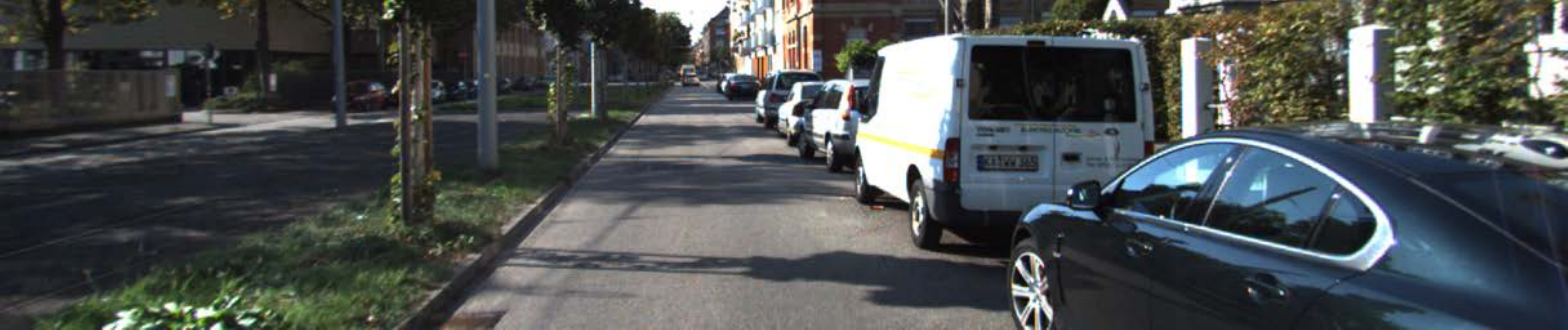}
		\caption{RGB image}
		\label{fig:channel-21}
	\end{subfigure} %
	\begin{subfigure}{0.24\linewidth}  
		\centering  
		\includegraphics[width=1.73in]{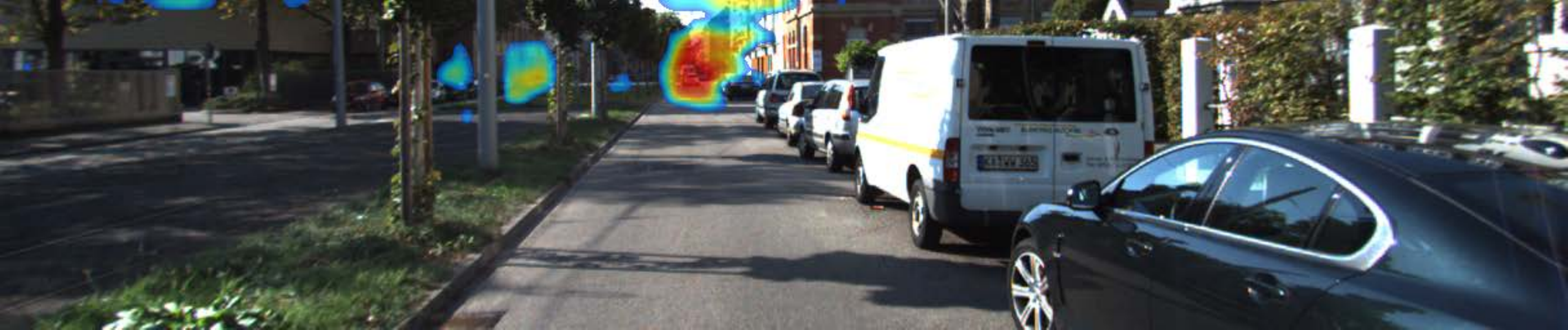}
		\caption{18th channel,$weight=0.946$}  
		\label{fig:channel-22}
	\end{subfigure} 
	\begin{subfigure}{0.24\linewidth}  
		\centering  
		\includegraphics[width=1.73in]{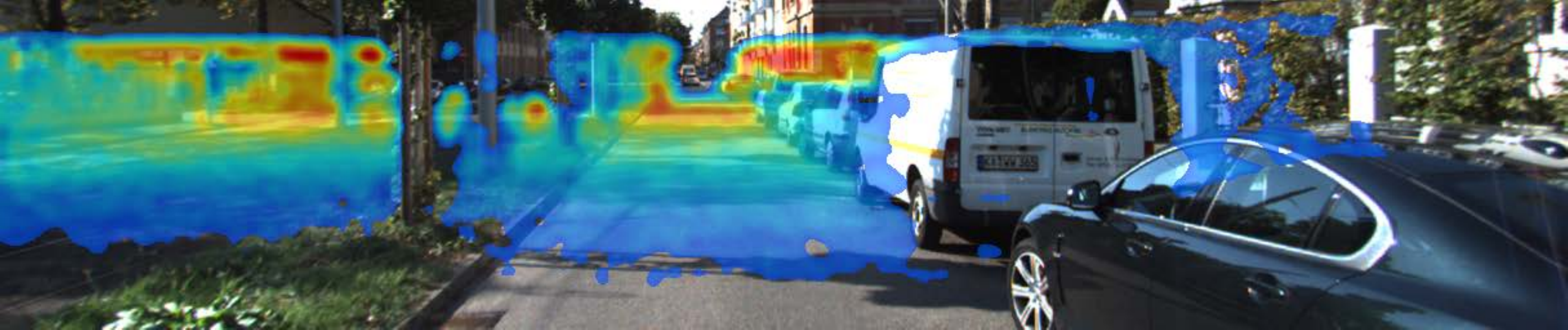}
		\caption{32th channel,$weight=0.999$}  
		\label{fig:channel-23}
	\end{subfigure}
	\begin{subfigure}{0.24\linewidth}  
		\centering  
		\includegraphics[width=1.73in]{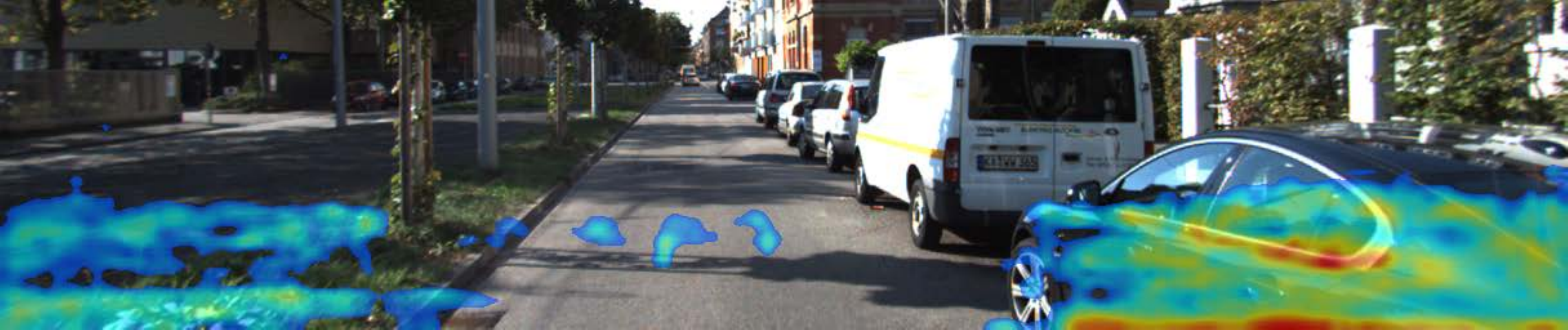}
		\caption{56th channel,$weight=0.527$}  
		\label{fig:channel-24}
	\end{subfigure}
	
	\begin{subfigure}{0.24\linewidth}
		\centering
		\includegraphics[width=1.73in]{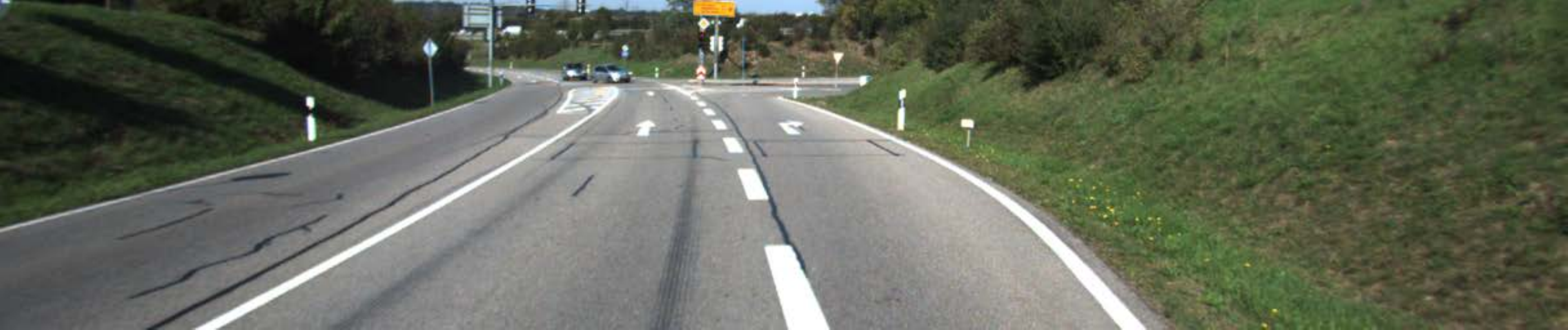}
		\caption{RGB image}
		\label{fig:channel-31}
	\end{subfigure} %
	\begin{subfigure}{0.24\linewidth}  
		\centering  
		\includegraphics[width=1.73in]{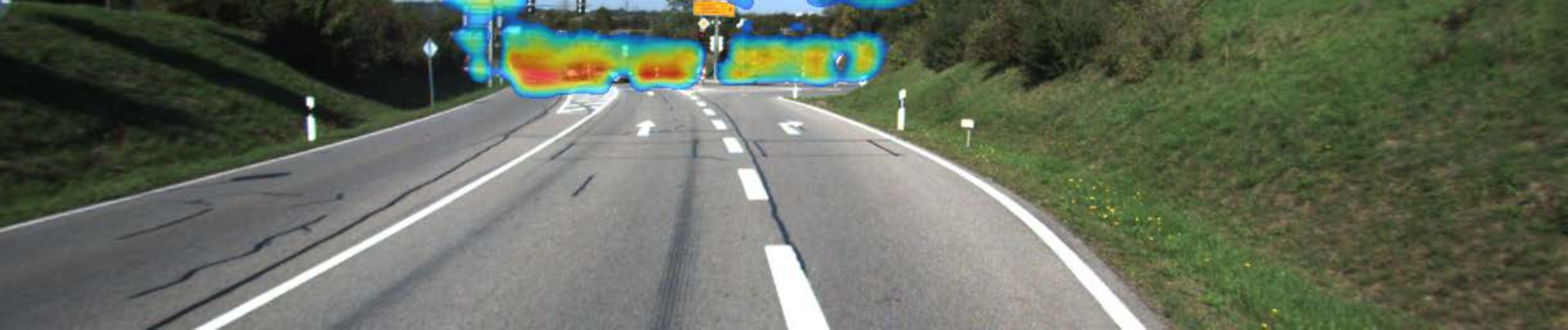}
		\caption{18th channel,$weight=0.930$}  
		\label{fig:channel-32}
	\end{subfigure} 
	\begin{subfigure}{0.24\linewidth}  
		\centering  
		\includegraphics[width=1.73in]{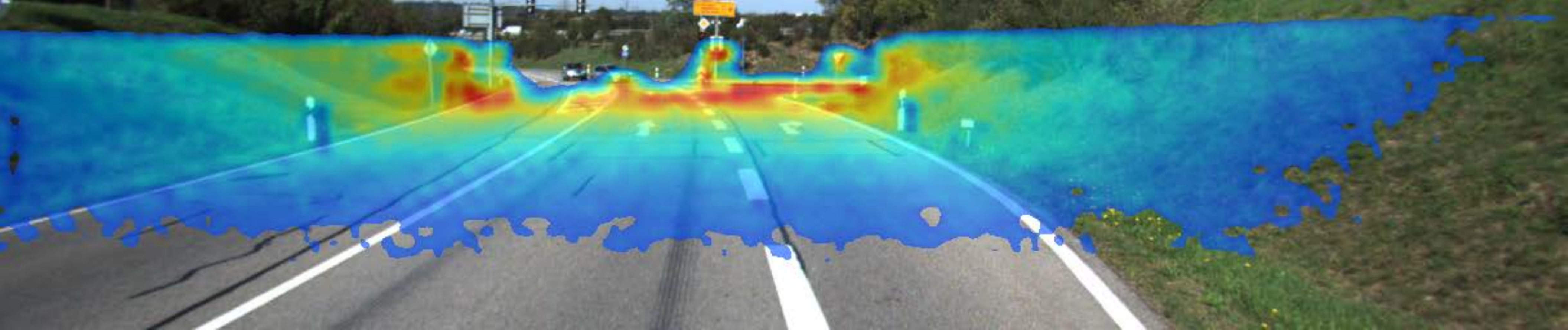}
		\caption{32th channel,$weight=0.999$} 
		\label{fig:channel-33}
	\end{subfigure}
	\begin{subfigure}{0.24\linewidth}  
		\centering  
		\includegraphics[width=1.73in]{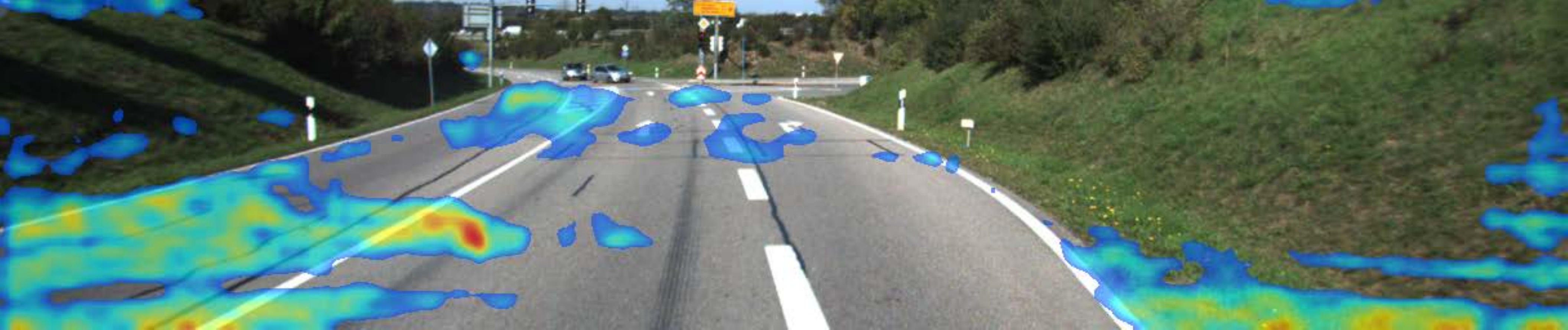}
		\caption{56th channel,$weight=0.514$} 
		\label{fig:channel-34}
	\end{subfigure}
	\caption{The activation map of features is converted from gray to the jet color map, and the converted color map is applied on RGB image. The first column is RGB images. The 2nd,3rd,4th columns are the visualization of 18th, 32nd,56th channels in feature map generated by encoder network. The 18th, 32th and 56th channel response to far objects, mid-distance objects and near objects. The weights of each channel generated by channel enhancer are shown in the caption of each picture.}
	\label{fig:channel}
	\setlength{\leftskip}{-10pt}
\end{figure*}
%%%%%%%%%%%%%%%%%%%%%%%%%%%%%%%%%%%%%%%%
\begin{table}[htbp]
	\caption{Comparisons to the state-of-the-art approaches on the KITTI on-line benchmark. For all metrics, lower is better. The best and the second best records are marked as \textbf{bold} and \color{blue}blue\color{black}.}
	\centering
	\scriptsize
	\setlength\tabcolsep{2pt} % default value: 6pt
	\label{tab:evaluation}
	\begin{threeparttable}[t]
		%\newcolumntype{M}{>{$\vcenter\bgroup\hbox\bgroup}c<{\egroup\egroup$}}
		\begin{tabular}{| l || *{6}{ c |}  }
			\hline
			\textbf{on KITTI benckmark} & \thead{iRMSE} & iMAE & RMSE & MAE & Runtime[ms] \\
			\hline \hline
			CSPN \etal~\cite{cheng2018depth}  & 2.93 & 1.15 & 1019.64 & 279.46 & 1000 \\
			DFineNet \etal~\cite{zhang2019dfinenet} & 3.21 & 1.39 & 943.89 & 304.17 & \color{blue} 20\\ 
			NConv-CNN-L1 \etal~\cite{eldesokey2018confidence} & 2.52 & \color{blue} 0.92 & 859.22 & \color{blue} 207.77 & \color{blue} 20 \\
			NConv-CNN-L2 \etal~\cite{eldesokey2018confidence} & 2.60 & 1.03 & 829.98 & 233.26 & \color{blue} 20 \\ 
			DDP \etal~\cite{yang2019dense} & \textbf{2.10} & \textbf{0.85} & 832.94 & \textbf{203.96} & 80 \\
			Sparse-to-Dense (gd) \etal~\cite{ma2018self-supervised} & 2.80 & 1.21 & 814.73 & 249.95 & 80 \\
			RGB guidececertainty \etal~\cite{van2019sparse} & 2.19 & 0.93 & 772.87 & 215.02 & \color{blue} 20 \\
			DeepLidar \etal~\cite{qiu2018deeplidar:} & 2.56 & 1.15 & \textbf{758.38} & 226.50 & 70\\
			\hline \hline
			This Work & \color{blue} 2.12 & \color{blue} 0.90 & \color{blue} 759.64 & 210.33 & \textbf{18}\\
			\hline
		\end{tabular}
	\end{threeparttable}
\end{table}
\subsection{Analysis of Spatial-wise Enhancer}
We utilize redesigned non-local operation as our spatial attention enhancer to enlarge the receptive field of our encoder network. To analyze the influence of our spatial enhancer. We visualize the attention maps produced by spatial enhancer, which are shown in Figure \ref{fig:spatial}. The first column is the raw RGB image. The second and third columns are the visualized attention of spatial enhancer. The corresponding point of the visualized attention map is marked with red dots. It can be found that the spatial enhancer can integrate features in a global view with their relationship regardless of their distance.
\subsection{Analysis of Channel-wise Enhancer}
The channel-wise enhancer is adapted to reassign the weights of channels. To analyze the influence of our channel enhancer module. We visualize the channels responding to different distance ranges, which are shown in Figure \ref{fig:channel}. The first column is the raw RGB image. The second column is the 18th channel, which response to far objects. The 3rd column is the 32nd channel, which response to middle-distance objects. The fourth column is the 56th channel response to near objects. The activation map of corresponding channels are visualized in the jet color map and applied on the RGB image. The weights generated by channel enhancer is written in the caption under the picture. It can be found that the channel enhancer can reassign the weight of channels from an attention-based mechanism. For example, as shown in the first row of Figure \ref{fig:channel}, this frame contains only a few far objects. Thus the weight of the 18th channel is significantly lower than the frame shown in the second and third row. This frame contains a lot of near objects. Thus the weight of the 56th channel is higher than the others. Meanwhile, all of the pictures mainly contain objects in the middle distance. Therefore, the 32nd channel is paid the highest attention.
\subsection{Comparision with Existing Methods}
\begin{figure*}[htbp] \centering
	\begin{subfigure}{0.24\linewidth}
		\centering
		\includegraphics[width=1.78in]{pics/compare_img}
		\caption{raw rgb image}
		\label{fig:compare-img}
	\end{subfigure} %
	\begin{subfigure}{0.24\linewidth}  
		\centering  
		\includegraphics[width=1.78in]{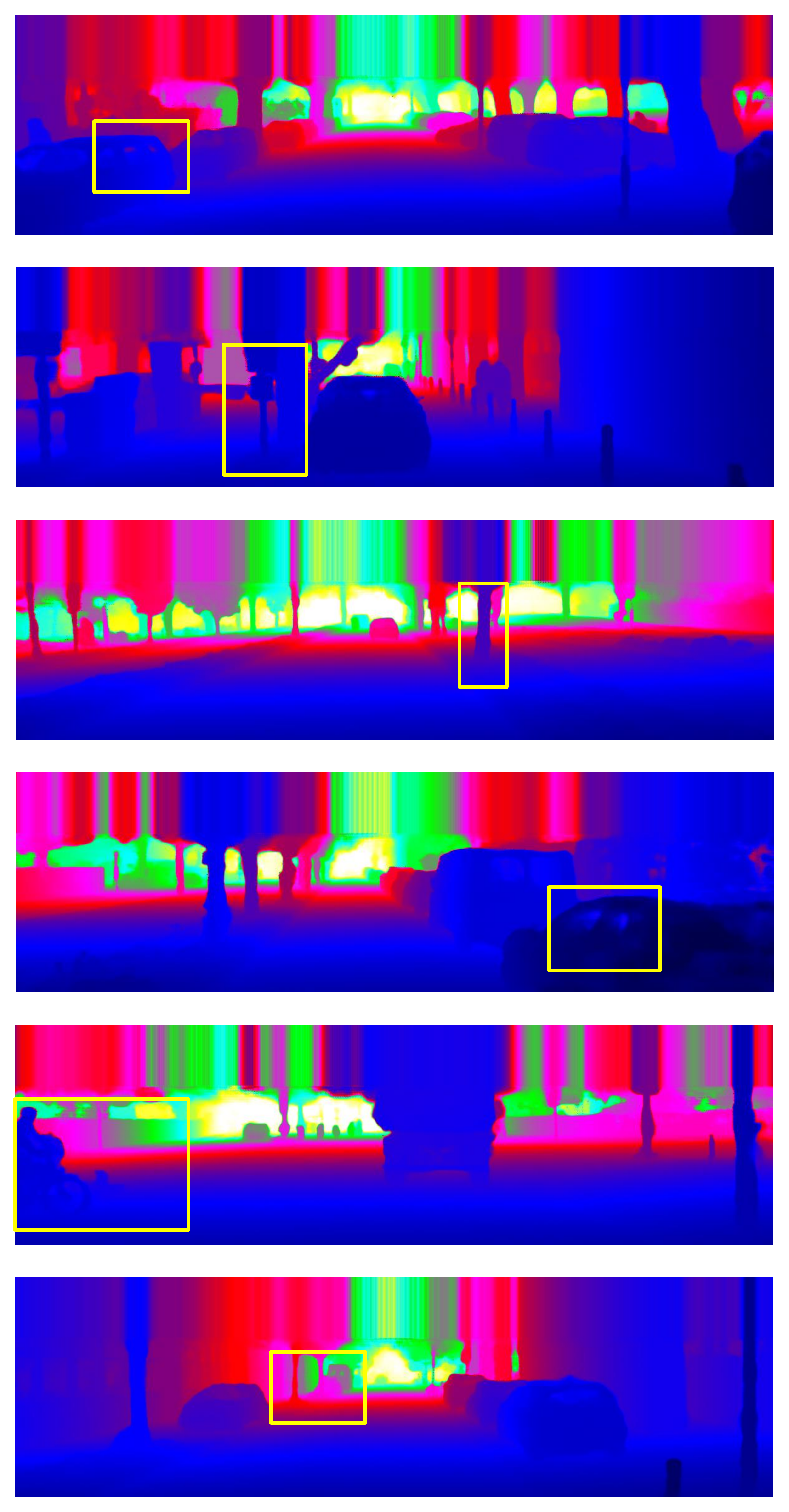}
		\caption{ours}  
		\label{fig:compare-ours}
	\end{subfigure} 
	\begin{subfigure}{0.24\linewidth}  
		\centering  
		\includegraphics[width=1.78in]{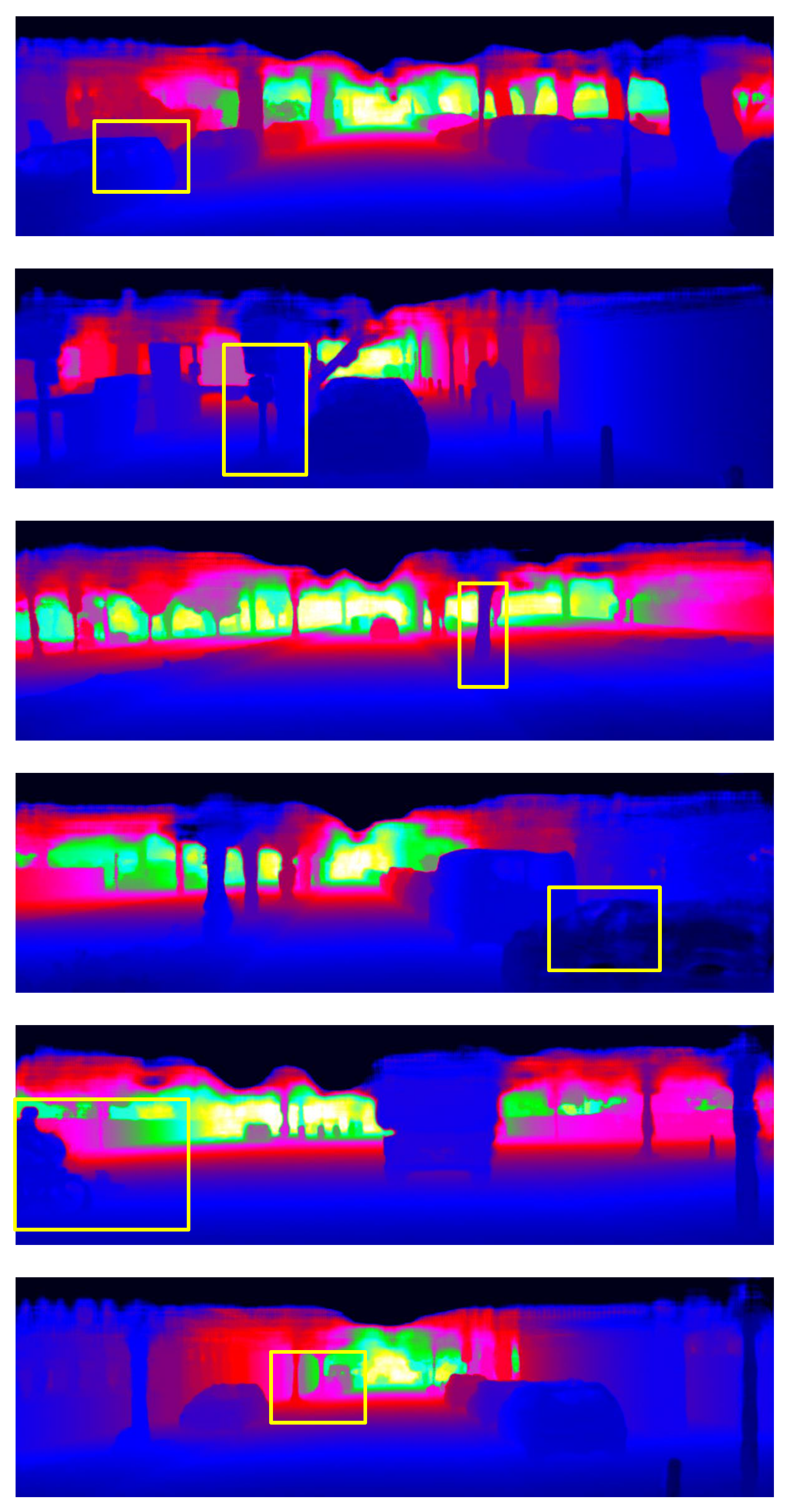}
		\caption{sparse-to-dense}
		\label{fig:compare-sparse}
	\end{subfigure} 
	\begin{subfigure}{0.24\linewidth}
		\centering  
		\includegraphics[width=1.78in]{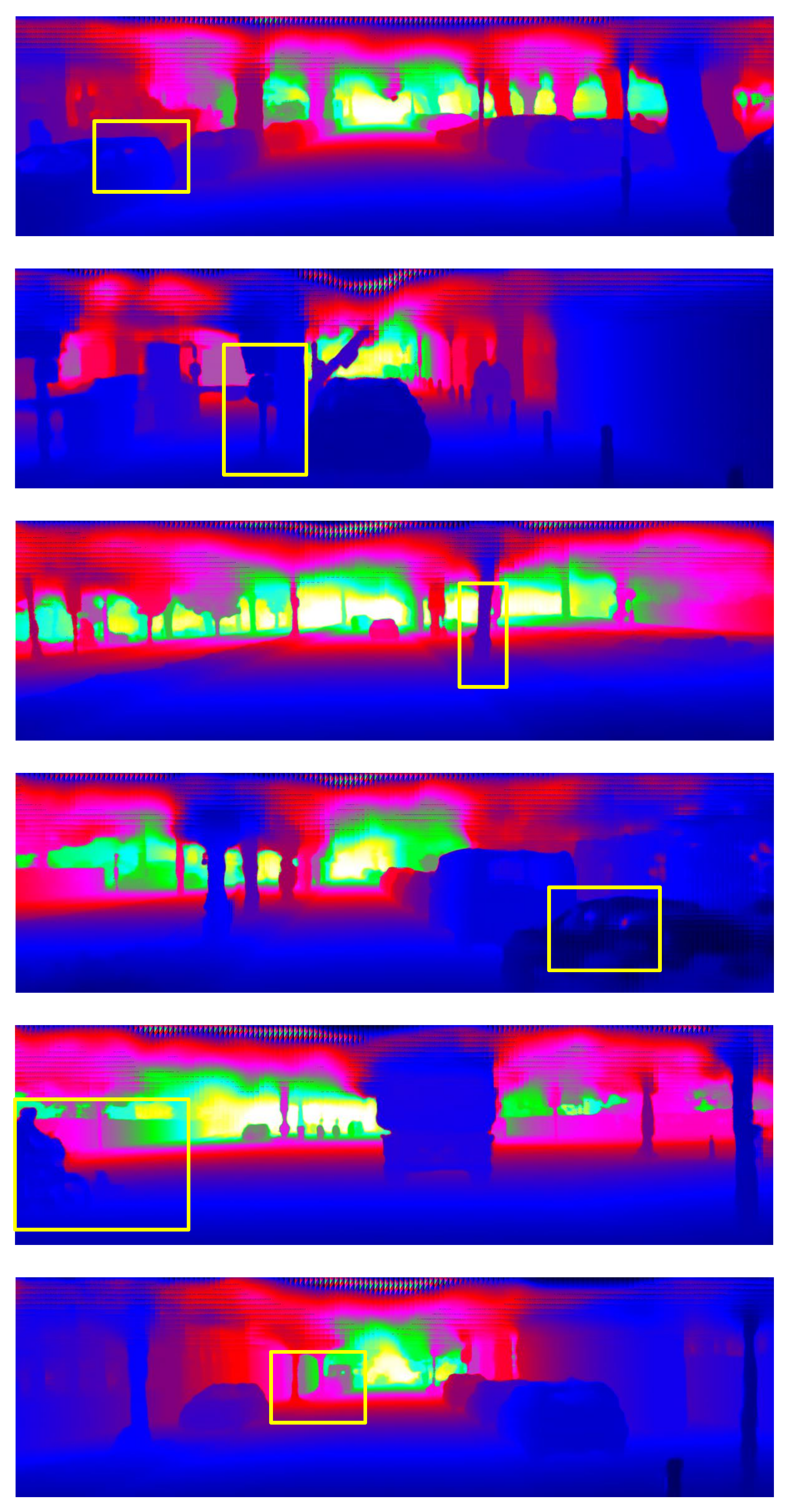}
		\caption{deeplidar}
		\label{fig:compare-deeplidar} 
	\end{subfigure} 
	
	\caption{Visual comparison with state-of-the-art. The yellow box shows the area to focus on in the depth map. Our approach shows sharper boundaries. For example, see the car's window in the first and third row.}
	\label{fig:compare-all}
\end{figure*}

In this section, we train our best network for KITTI dataset against other published approaches. We use the official error metrics on the KITTI depth completion benchmark. The results are listed in Table \ref{tab:evaluation} and visualized in Figure \ref{fig:compare-all}.

As shown in Table \ref{tab:evaluation}, our approach achieves comparable performance with the published state-of-the-art in the ranking metric (\textsl{RMSE}) but outperform it in all other metrics with a much higher inference speed. The DDP in \cite{yang2019dense} performs best in terms of metrics \textsl{iRMSE},\textsl{iMAE} and \textsl{MAE} as they employed \textsl{MAE} as their loss function. However, their performance on the ranking metric (\textsl{RMSE}) is much worse than other state-of-the-art approaches. The approaches in \cite{zhang2019dfinenet,eldesokey2018confidence} run much faster than other state-of-the-art methods, but their performance on metrics are much worse. Both the proposed method and the method in \cite{van2019sparse} achieve a good balance between accuracy and speed. However, our approach outperforms the method in \cite{van2019sparse} in all metrics with a similar running speed. 

Moreover, our estimated depth image also have a cleaner and sharper object boundaries even compares with the state-of-the-art methods, which can be attributed to the fact that our S\&C model boosts the representational power of our encoder network and the lower output stride of backbone network can preserve more image details. For example, the car's window in left-bottom of the first picture and the car's window in right-bottom of the third picture. Note that we trained our network with an input size of $1216 \times 256$ because the $30\%$ semi-dense annotations do not contain labels in these top regions. At the evaluation, we just expand the image into the resolution of $1216 \times 352$ by copy the top row of the original output depth map.
%%%%%%%%%%%%%%%%%%%%%%%%%%%%%%%%%%%%%%%%%%%%%%%%
%%%%%%%%%%%%%%%%%%%%%%%%%%%%%%%%%%%%%%%%%%%%%%%%
\begin{table}[thbp]
	\caption{The compare result of original model, finetuned model and finetuned model after plugging the proposed S\&C enhancer. }
	\centering
	\scriptsize
	\setlength\tabcolsep{3pt} % default value: 6pt
	\label{tab:compare}
	\begin{subtable}{.5\textwidth}
		\centering
		\caption{The compare result of work sparse-to-dense \cite{ma2018self-supervised}.}
		\begin{tabular}{| c || *{3}{ c } | }
			\hline
			\textbf{Model} & \textbf{RMSE} & \textbf{Parameters (Million)} & \textbf{Inference time} \\
			Original & 858 & 26.107 & 78.6\\
			Finetuned & 855 & 26.107 & 78.6\\
			Enhanced & 806 & 26.485 &79.9\\
			\hline
		\end{tabular}
	\end{subtable}%
	\vspace{0.3cm}
	\begin{subtable}{.5\textwidth}
		\centering
		\caption{The compare result of work RGB-guideconfidence \cite{van2019sparse}.}
		\begin{tabular}{| c || *{3}{ c } | }
			\hline
			\textbf{Model} & \textbf{RMSE} & \textbf{Parameters (Million)} & \textbf{Inference time (ms)} \\
			Original & 802 & 2.545 & 19.8\\
			Fineuned & 804 & 2.545 & 19.8\\
			Enhanced & 791 & 2.568 & 20.4\\
			\hline
		\end{tabular}
	\end{subtable}%
\end{table}
%%%%%%%%%%%%%%%%%%%%%%%%%%%%%%%%%%%%%%%%%%%%%%%
%%%%%%%%%%%%%%%%%%%%%%%%%%%%%%%%%%%%%%%%%%%%%%%%%
\begin{table}[htbp]
	\caption{Comparisons of sparse-to-dense and enhanced sparse-to-dense on the on-line KITTI benchmark.}
	\centering
	% \small
	% \footnotesize
	\scriptsize
	\setlength\tabcolsep{3pt} % default value: 6pt
	\label{tab:sparse-to-dense}
	\begin{threeparttable}[t]
		%\newcolumntype{M}{>{$\vcenter\bgroup\hbox\bgroup}c<{\egroup\egroup$}}
		\begin{tabular}{| l || *{6}{ c |}  }
			\hline
			\textbf{on KITTI benckmark} & \thead{iRMSE} & iMAE & RMSE & MAE & Runtime[ms] \\
			\hline \hline
			Original  & 2.60 & 1.21 & 814.73 & 249.95 & 80 \\
			%\hline \hline
			Enhanced & \textbf{2.40} & \textbf{1.08} & \textbf{772.66} & \textbf{231.89} & 80\\
			\hline
		\end{tabular}
	\end{threeparttable}
\end{table}
%%%%%%%%%%%%%%%%%%%%%%%%%%%%%%%%%%%%%%%%%
\subsection{Boosting Power of the S\&C Enhancer}
The previous experimental result has demonstrated the improvement of our proposed S\&C enhancer in our network. Meanwhile, our proposed S\&C enhancer can be easily plugged into existing deep architectures and produce significant performance improvements with marginal additional computational cost. To demonstrate the boosting power of our proposed S\&C enhancer, we investigate the effect of plugging S\&C enhancer into other two existing architectures, sparse-to-dense \cite{ma2018self-supervised} and confidence-guidance \cite{van2019sparse}. We first download the pre-trained model of these two networks and implement two ablation studies over these two networks. We fine-tuned the two pre-trained models directly for ten epochs with the default training strategies at first. As a comparison, we plug the proposed S\&C enhancer at the end of the encoder in the two architectures and fine-tuned the enhanced models for ten epochs from the download parameters. We evaluate the results of both networks in the KITTI selected cropped validation dataset. Besides, we submit the testing result of enhancer sparse-to-dense \cite{ma2018self-supervised} to the KITTI on-line benchmark.

The finetuned result in KITTI selected cropped validation dataset is listed in Table \ref{tab:compare}. In the column \textbf{Model}, \textsl{"Original"} stand for the performance of the downloaded model, \textsl{"Finetuned"} stands for the fine-tuned model without proposed S\&C enhancer and \textsl{"Enhanced"} stand for the fine-tuned model after plugged the proposed S\&C enhancer. For the sparse-to-dense, the proposed S\&C enhancer improved the RMSE from $858mm$ to $806mm$ with only 0.378 million additional parameters. For the RGB-guideconfidence, the proposed S\&C enhancer improved the RMSE from $802mm$ to $791mm$ with only 0.023 million additional parameters.

We submit the result of enhanced sparse-to-dense to KITTI on-line benchmark and the performance is improved significantly. The compare result of metrics is listed in Table \ref{tab:sparse-to-dense} and the compare result is shown in Figure \ref{fig:compare-ours}. For the sparse-to-dense, our S\&C enhancer improves RMSE from $814.73mm$ to $772.66mm$. With the plugged S\&C enhancer, the ranking of sparse-to-dense is improved from 19th to 9th.

\section{Conclusion} 
\label{sec:conclusion}
We have proposed a novel coarse-to-fine network for the dense depth completion for real-time requirements of autonomous systems and 3D reconstruction. The proposed method achieves competitive accuracy performance with the state-of-the-art approaches but almost $4$ times faster. In addition, our proposed S\&C enhancer can be easily plugged into other existing networks and boost their performance on the dense depth completion significantly with a small amount of additional parameters. Our current study focuses on the case that the input image is clean. It is also necessary to study the case that the input image is distorted such as a haze image \cite{li2017single}. This problem will be studied in our future research.

%\clearpage
\bibliographystyle{IEEEtranN}
\nocite{*}
%\bibliography{reference}
\input{scnet.bbl}

\end{document}

%% file: scnet.bbl
% Generated by IEEEtranN.bst, version: 1.13 (2008/09/30)